\documentclass[manuscript]{emulateapj}
\usepackage{textcomp}
\usepackage{amssymb}
\usepackage[normalem]{ulem}
\usepackage{amsmath}
\usepackage{graphicx}

\shorttitle{Old stellar substructures in MW's dSphs}
\shortauthors{Lora et al.}

\begin{document}

\title{Cold, old and metal-poor: New stellar substructures \\
in the Milky Way's dwarf spheroidals}

\author{V. Lora\altaffilmark{1,2}, E. K. Grebel\altaffilmark{2}, S. Schmeja\altaffilmark{2,3},
 and A. Koch\altaffilmark{2}}
\affil{\altaffilmark{1}Instituto de Radioastronom\'ia y Astrof\'isica, \\
             Antigua Carretera a P\'atzcuaro 8701, 58089 Morelia, M\'exico}
\affil{\altaffilmark{2} Astronomisches Rechen-Institut, Zentrum f\"{u}r Astronomie der Universit\"{a}t Heidelberg, \\
             M\"{o}nchhofstr. 12-14, 69120 Heidelberg, Germany}
\affil{\altaffilmark{3} Technische Informationsbibliothek, \\
	     Welfengarten 1b, 30167 Hannover, Germany}           

\email{v.lora@irya.unam.mx}

\begin{abstract}
Dwarf spheroidal galaxies (dSph) orbiting the Milky Way are complex objects often with 
complicated star formation histories and internal dynamics. In 
this work, we search for stellar substructures in four of the classical dSph satellites 
of the Milky Way: Sextans, Carina, Leo~I, and Leo~II. We apply two methods to search for 
stellar substructure: the minimum spanning tree method, which helps us to find and quantify 
spatially connected structures, and the ``brute-force'' method, which is able to find 
elongated stellar substructures. We detected the previously known substructure in Sextans, 
and also found a new stellar substructure within Sextans. Furthermore, we identified a new 
stellar substructure close to the core radius of the Carina dwarf galaxy. We report a detection
of one substructure in Leo~I and two in  Leo~II, but we note that we are dealing with a low 
number of stars in the samples used. Such old stellar substructures in dSph galaxies could 
help us to shed light on the nature of the dark matter halos, within which such structures 
form, evolve, and survive.

\end{abstract}



\section{Introduction}
Dwarf spheroidal galaxies (dSph) are the most common type of galaxies in the 
Universe and often considered to be the building blocks of more massive galaxies in hierarchical 
formation scenarios \citep{dekel:86,bullock:05,cooper:13,pillepich:15}. The dSph satellites of 
the Milky Way (MW) are the best studied dwarf galaxies, since individual stars can be resolved and 
evolutionary histories can be derived in great detail. A common property of all dSph galaxies 
is the presence of an old population (stars with ages of $10$~Gyr and more), which in many cases 
turns out to be the dominant population e.g. \citep{grebel:00, grebel:04}.
There is growing evidence, that some dSphs (e.g., Carina, Fornax, Sculptor, and Sextans) 
experienced extended star formation histories or multiple episodes of star formation
The generally spatially extended old, metal-poor population, 
and spatially concentrated young, metal-rich population indicate extended star formation episodes
in their centers \citep{harbeck:01}. 

Some dSph galaxies also present dynamically cold stellar substructures. For example, Ursa Minor 
(UMi) shows two distinct density peaks \citep{kleyna:98}. One represents the underlying galaxy's 
field population with a radial velocity dispersion of $\sigma=8.8$~km~s$^{-1}$. The second 
density peak appears with a radial velocity dispersion of $\sigma=0.5$~km~s$^{-1}$.
This second peak is located on the north-eastern side of the major axis of UMi at a distance of 
$\sim 0.4$~kpc from UMi's center. The stars in the vicinity of this peak comprise a kinematically 
distinct cold sub-population, possibly a disrupted cluster, i.e., a dynamically cold stellar clump.

These clumpy substructures indicate that there is a high degree of complexity already on small
scales, which it is also witnessed in other systems (such as in the distribution of the globular 
clusters in the Fornax dSph), mergers between dwarf galaxies in the local Universe 
\citep{amorisco:14,delpino:15} 
and a possible merger event in the Large Magellanic Cloud (LMC) \citep{mackey:16}.

Another example is found in Sextans. \cite{kleyna:04} reported the line-of-sight radial velocity 
dispersion profile of Sextans based on a sample of $88$ stars extending to about $1.5$~kpc from 
the center of Sextans. They found that the dispersion at the center of Sextans was close to zero, 
and that such a low value of the dispersion was in agreement with significant radial gradients in 
the stellar populations. They suggested that this is caused by the sinking and gradual dissolution 
of a star cluster at the center of Sextans. On the other hand, \cite{walker:06} presented $294$ 
possible stellar Sextans members. Their data did not confirm \citeauthor{kleyna:04}'s (\citeyear{kleyna:04}) 
report of a kinematically distinct stellar population at the center of Sextans with their more 
complete sample, but they obtained similar evidence if they restricted their analysis to a similarly small number 
of stars as used by \cite{kleyna:04}. Instead, \cite{walker:06} detected a region near Sextans' core 
radius, which is kinematically colder than the overall Sextans sample with 95\% confidence. Very 
recently, \cite{kim:19} reported the ``possible detection of a relic'' globular cluster in Sextans.

Another example is provided by the possibly distinct stellar populations in the Canes Venatici I (CVn~I) dSph 
\citep{zucker:06}, where \cite{ibata:06} found two stellar populations with different kinematics: an 
extended metal-poor ([Fe/H]$<-2$) component with a velocity dispersion of $\sim10$~km~s$^{-1}$, 
and a more concentrated metal-rich component of extremely low velocity dispersion. On the other 
hand, \cite{simon:07} obtained $214$ CVn I member stars that did not reveal any trace of different 
populations as reported by \cite{ibata:06}. 
\cite{ural:10} obtained spectroscopic data for $26$ stars in the dSph and 
investigated whether their data exhibit any evidence of multiple populations as proposed by 
\cite{ibata:06} (under the assumption that each population was Gaussian), but no clear signature 
of distinct sub-populations was found. They argue that the possible detection of a sub-population 
rather depends on the total number of stars in the data set, the fraction of stars in the sub-population, 
the difference in velocity dispersion between the populations, and the observational errors.

A final example is the Andromeda~II (And~II) dSph. \cite{mcconnachie:07} identified different 
stellar populations in And~II for which they constructed radial profiles and found that the 
horizontal branch has a constant spatial distribution out to a large radius. In contrast, they 
found that the reddest stars on the red giant branch in And~II are more centrally concentrated. 
The latter stellar component has an average age of $\sim 9$~Gyr and is relative metal-rich 
([Fe/H]$\sim -1$), leading to a prominent red clump. And, it has to be noted that And~II has 
been identified as having experienced a merger \citep{amorisco:14}.

The stellar substructures found within dSphs are of great importance, since the survival of 
stellar substructures within the dark matter (DM) halo of dSphs suggests that dSphs are more in 
agreement with a cored DM profile, rather than with the cuspy DM profile 
predicted by the $\Lambda$-cold DM model \citep{lora:12,lora:13,amorisco:17,contenta:17}.

Using N-body simulations, \cite{lora:12} and \cite{lora:13} studied the survival of
cold kinematic stellar substructures embedded in the DM halo of the Fornax and Sextans
dwarf galaxies -respectively-, against phase mixing. They compared the evolution of the
stellar substructures when the dark halo has a core and when it follows the 
Navarro-Frenk-White (NFW) profile \citep{nfw}. The core in the DM halo mass density 
distributions is large enough to make the gravitational potential almost harmonic, 
guaranteeing the survival of the stellar structures mentioned even if the stellar 
substructure is initially very extended. On the contrary, the stellar substructures 
are very rapidly destroyed when embedded in a NFW DM profile.

Still, one cannot rule out a scenario where the DM profile of dwarf galaxies was 
initially cuspy and evolved to a cored DM halo via energy feedback from supernova
explosions, stellar winds, and/or star formation \citep{mashchenko:06,chen:16,pasetto:10}.
However, in faint galaxies like Eridanus~II and Andromeda~XXV 
(i.e., $L_{V}\sim5\times10^{5}$~L$_{\odot}$ and $L_{V}\sim6\times10^{4}$~L$_{\odot}$, respectively)
the formation of large cores via stellar feedback is not obviously expected \citep{amorisco:17}.
If stellar feedback in faint galaxies is found to be inefficient, then alternative 
candidates for DM should be seriously considered. 

The detection of stellar substructures in dSphs, and the implications of their survival in 
the core/cusp problem, prompted us to test other dwarf galaxies, and to investigate if we 
could find stellar substructures within them. From velocities, metallicities, and positions 
of red giant stars in the Sextans, Carina, Leo I, and Leo II dSphs, we analyze whether or not 
these dSphs could host dynamically cold debris.

This article is organized as follows: In \S\ref{sec:dwarfs} we describe some properties of Sextans, 
Carina, Leo~I, and Leo~II. We describe the methods used to search for substructures in 
\S\ref{sec:methods}. The results are presented in \S\ref{sec:results}. Finally, we present our 
conclusions in \S\ref{sec:conclusions}. 



\section{The four dwarf spheroidal galaxies}
\label{sec:dwarfs}
\begin{table*}
\begin{center}
\caption{Parameters of Sextans, Carina, Leo~I and Leo~II dSphs.}
 \begin{tabular}{@{}ccccccc@{}}
\tableline
& & & & &\\
Parameter & Sextans &Carina & Leo~I & Leo~II &Ref.$^{a}$ &\\
\tableline
& & & & &\\
($\alpha_{J2000},\delta_{J2000}$)&10$^{h}$13$^{m}$03$^{s}$, -01\textdegree13$^{'}$03${''}$& 6$^{h}$40$^{m}$24.3$^{s}$, -50\textdegree58$^{'}$0${''}$&10$^{h}$8$^{m}$27.0$^{s}$, 12\textdegree18$^{'}$30${''}$ & 11$^{h}$13$^{m}$29$^{s}$, 22\textdegree9$^{'}$12${''}$ & 1,1,1,1 &\\
Distance     & $86\pm6$~kpc&$105\pm6$~kpc &$254\substack{+16 \\ -19}$\,kpc    &  $233\pm15$~kpc  & 2,2,2,2 & \\
M$_{V}$      & $-9.3\pm0.5$~mag    &$-9.3\pm0.5$~mag     &-11.9~mag         & -9.6~mag  &2,2,3,2 & \\
PA           & 57          & 60          &78                & 12        &11,11,11,5 & \\
$e$          & 0.30        & 0.38        &0.31              & 0.07      &11,11,11,11 & \\
R$_{core}$   & 20.1$\pm0.05$~arcmin       & 7.97$\pm0.16$~arcmin   &3.6$\pm0.1$~arcmin   & 2.25$\pm0.1$~arcmin & 11,11,11,11 & \\
R$_{tidal}$  & 60.5$\pm0.6$~arcmin         & 58.4$\pm0.98$~arcmin  &13.5$\pm0.3$~arcmin  & 9.82$\pm0.4$~arcmin & 11,11,11,11 & \\
v$_{sys}$    & 226 $\pm0.6$~km~s$^{-1}$  & 223.9~km~s$^{-1}$    &284.2~km~s$^{-1}$    & 79.1~km~s$^{-1}$  & 6,7,8,9 & \\
M/L          & 97          &  66         &24                & 25-50     & 10,10,8,9 &\\
\textlangle[Fe/H]\textrangle& -1.93$\pm 0.1$ & -1.72$\pm 0.1$& -1.43$\pm 0.1$ & -1.62$\pm 0.1$ & 6,7,8,9 &\\
\tableline
\end{tabular}
\label{tab:carina}
\footnotetext{References: (1) \cite{mateo:98}, (2) \cite{mcconnachie:12}, (3) \cite{bellazzini:04},
(4) \cite{bellazzini:05}
(5) \cite{irwin:95}, (6) \cite{battaglia:11}, (7) \cite{koch:06}, (8) \cite{koch_leoI:07}, 
(9) \cite{koch_leoII:07}, \cite{lokas:09} \& (11) \cite{munoz:18}.}
\end{center}
\label{table:parameters}
\end{table*}
\subsection{Sextans}
\label{sec:sextans}
The Sextans dSph is located at a distance of $86\pm6$~kpc
\citep{mcconnachie:12}. It is a diffuse and faint 
dSph, with $M_{V}=-9.3 \pm 0.5$~mag and  $\mu_{V}=27.1 \pm 0.5$~mag/sq~arcsec 
\citep{mcconnachie:12}. It has a position angle $PA=57$\textdegree and an 
ellipticity $e=0.30$ \citep{munoz:18}. \cite {battaglia:11} report the 
systemic velocity in the heliocentric system of Sextans to be 
$v_{sys}=226 \pm 0.6$~km~s$^{-1}$.

\cite{irwin:95} computed a core radius $r_{c}=16.6\pm1.2$~arcmin 
and a tidal radius $r_{t}=160\pm50$~arcmin. More recently, \cite{roderick:16} 
fitted a King model to the radial distribution of Sextans computing 
a core radius $r_{c}=26.8\pm1.2$~arcmin and a tidal radius 
$r_{t}= 83.2\pm7.1$~arcmin ($2.08\pm0.18$~kpc)
\cite{munoz:18} reported a core radius $r_{c}=20.1\pm0.05$~arcmin 
and a tidal radius $r_{t}=60.5\pm0.6$~arcmin 
(see also \citeauthor{cicuendez:18}, \citeyear{cicuendez:18}).

The Sextans dSph is more than $12$~Gyr old \citep{mateo:91} and it is considered 
as a metal poor dwarf galaxy \citep{kirby:01} with a mean [Fe/H]$=-1.93$ 
\citep{battaglia:11}. \cite{lokas:09} computed a high mass-to-light-ratio of $M/L \approx 97$.
\cite{battaglia:11} computed a $M/L \approx 460-920$.
Such a high mass-to-light-ratio indicates that Sextans is a DM-dominated dwarf galaxy.

The existence of stellar substructure in the Sextans dwarf galaxy has been 
previously studied. For example, \cite{kleyna:04} found a drop in the 
stellar velocity dispersion near the center of Sextans, which was interpreted 
as a dissolving cluster. Some years later, 
\cite{walker:06} reported no detection of a kinematically distinct 
population at the center of Sextans. Instead, they detected a region 
at a distance $r\sim16$~arcmin ($400$~pc) from Sextans' center, corresponding
to the core radius reported by \cite{irwin:95}.

Later, \cite{battaglia:11} reported the detection of a cold 
substructure consisting of nine very metal-poor stars close to the 
center of Sextans.These stars have very similar distances, kinematics, 
and metallicities. The average metallicity of this 9-star group is 
[Fe/H]$=-2.6$~dex with a $0.15$ dex $1\sigma$-scatter. This group of 
stars was taken from the most metal-poor stars, which have a low velocity 
dispersion ($\sim1.2$~km~s$^{-1}$) and whose average radial velocity is 
$72.5 \pm 1.3$~km~s$^{-1}$. \cite{battaglia:11} assume that the ratio 
of the stars in the substructure is representative with respect to 
their total number of Sextans members ($174$ stars). Thus 
the substructure would account for $5$\% of Sextans stellar population.

Finally, \cite{roderick:16} reported on photometric evidence of stellar 
substructures associated with Sextans. The stellar substructures that 
they find extend out to a distance of $82$~arcmin ($2$~kpc) from Sextans' 
centre. The existence of such stellar substructures in the outer regions 
of Sextans might indicate that Sextans is undergoing tidal disruption. 
However, \cite{roderick:16} found that the substructures surrounding Sextans 
appear to be both aligned with, and perpendicular to Sextans' major axis.
The latter suggests that Sextans is not necessarily undergoing a strong
tidal disruption. \cite{munoz:18} reported a fairly regular morphology in
Sextans, with no obvious signs of tidal features.

In contrast, recently \cite{cicuendez:18} reported signs of past accretion/merger 
events in Sextans: a `ring-like' feature. They claim that the  kinematically  
detected  ring in Sextans bears a morphological resemblance to the stellar 
stream in the And II dSph \citep{amorisco:14}, which probably merged with
another dwarf galaxy.

\subsection{Carina}
\label{sec:carina}                          
Carina is located at a heliocentric distance of $105\pm6$~kpc, with 
$M_{V}=-9.3 \pm0.5$~mag and  $\mu_{V}=25.5\pm0.5$~mag/sq~arcsec \citep{mcconnachie:12}.
It has a position angle $PA=60$\textdegree, a core radius 
$r_{c}=7.97\pm0.16$~arcmin, a tidal radius $r_{t}=58.4\pm0.98$~arcmin, 
an ellipticity $e=0.38$ \citep{munoz:18}, 
and a heliocentric systemic velocity $v_{sys}=223.9$~km~s$^{-1}$ \citep{koch:06}. 

The large velocity dispersion of Carina is often interpreted as evidence of a large 
mass-to-light ratio \citep{mateo:93}. \cite{lokas:09} found a mass-to-light ratio for
Carina of $66$, from where they inferred that Carina's kinematics 
is dominated by its DM halo. 

Spectroscopic observations of Carina indicate that it has two dominant stellar
populations that should be in equilibrium in the same DM halo, and that such a DM
halo has a less cuspy inner density profile than previously thought \citep{hayashi:18}.

The Carina dSph has experienced a complex star formation history. It is the only 
dSph to exhibit clearly episodic star formation interrupted by long quiescent 
periods \citep{deboer:14,santana:16,hurley:98,monelli:03}. Its color-magnitude diagram shows 
three different stellar populations corresponding to  $11$, $5$, and $0.6$~Gyr 
\citep{monelli:03, bono:10} and shows a slight population gradient where the more 
metal rich population is more centrally concentrated in the galaxy \citep{koch:06}.
\cite{deboer:14} confirmed that the star formation in the Carina dSph is episodic.
There are two main episodes of star formation that occurred at $>8$~Gyr and $2-8$~Gyr
(or at $\sim12$~Gyr and $4-8$~Gyr according to \citeauthor{monelli:14}, \citeyear{monelli:14}). 
\cite{monelli:14} argue that these two episodes of star formation are inconsistent with a simple 
evolution of an isolated system.

\cite{kordopatis:16} found that the youngest metal-rich population in Carina is more 
extended than the intermediate-metallicity population, while generally the metal-rich 
stellar populations are more spatially concentrated and kinematically colder. To explain 
this, \cite{sales:10} argue that strong tidal interaction may play a big role on 
Carina's peculiar configuration. In addition to this scenario, \cite{munoz:06} detected 
an extended power-law component in the density distribution of Carina. Such a distribution 
is a typical signature of disrupted galaxy satellites (e.g. the Sagittarius dSph, 
\citeauthor{majewski:03} \citeyear{majewski:03}). Moreover, \cite{munoz:06} found that
the extra-tidal stars around Carina could be related to the LMC, indicating complex
interactions.

\cite{fabrizio:11} presented radial velocity measurements of Carina's stars and collected 
spectra of old, intermediate, and young stellar tracers. They detected a maximum in the 
radial velocity, which they suggest might  be  reminiscent of a stellar 
structure located at $\sim200$~pc of Carina's center.
\cite{fabrizio:16} compared observations of the radial velocity distribution of old and
intermediate-age stars in the Carina dSph with N-body simulations. They found a good agreement
with the $V_{rot}/\sigma$ ratio in the central regions of the dwarf. The latter indicates 
that Carina might have been a disky dwarf galaxy that experienced several strong tidal 
interactions with the MW.
%
%
\begin{figure*}
\centering
(a){{\includegraphics[width = 7cm]{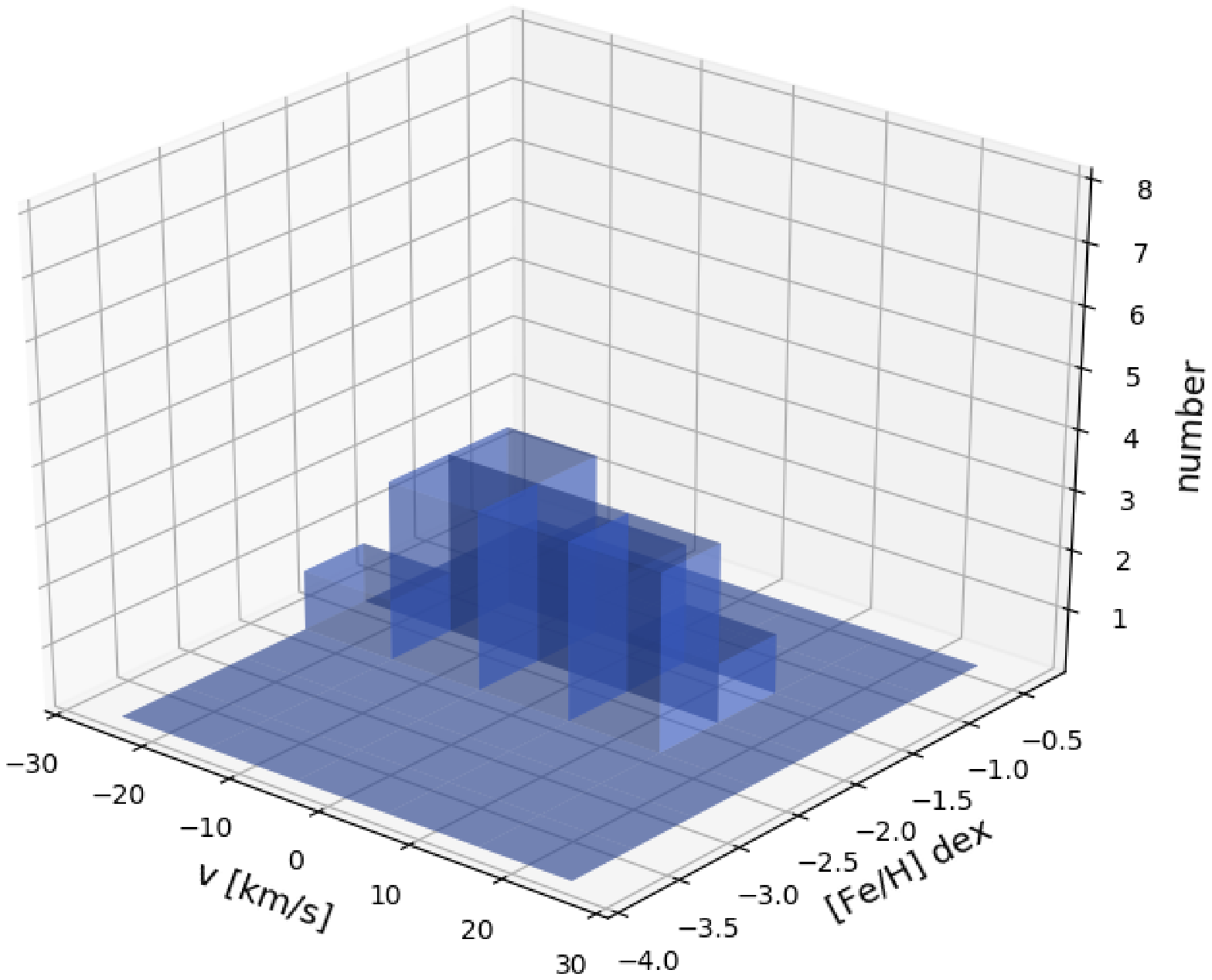}}}
(b){{\includegraphics[width = 7cm]{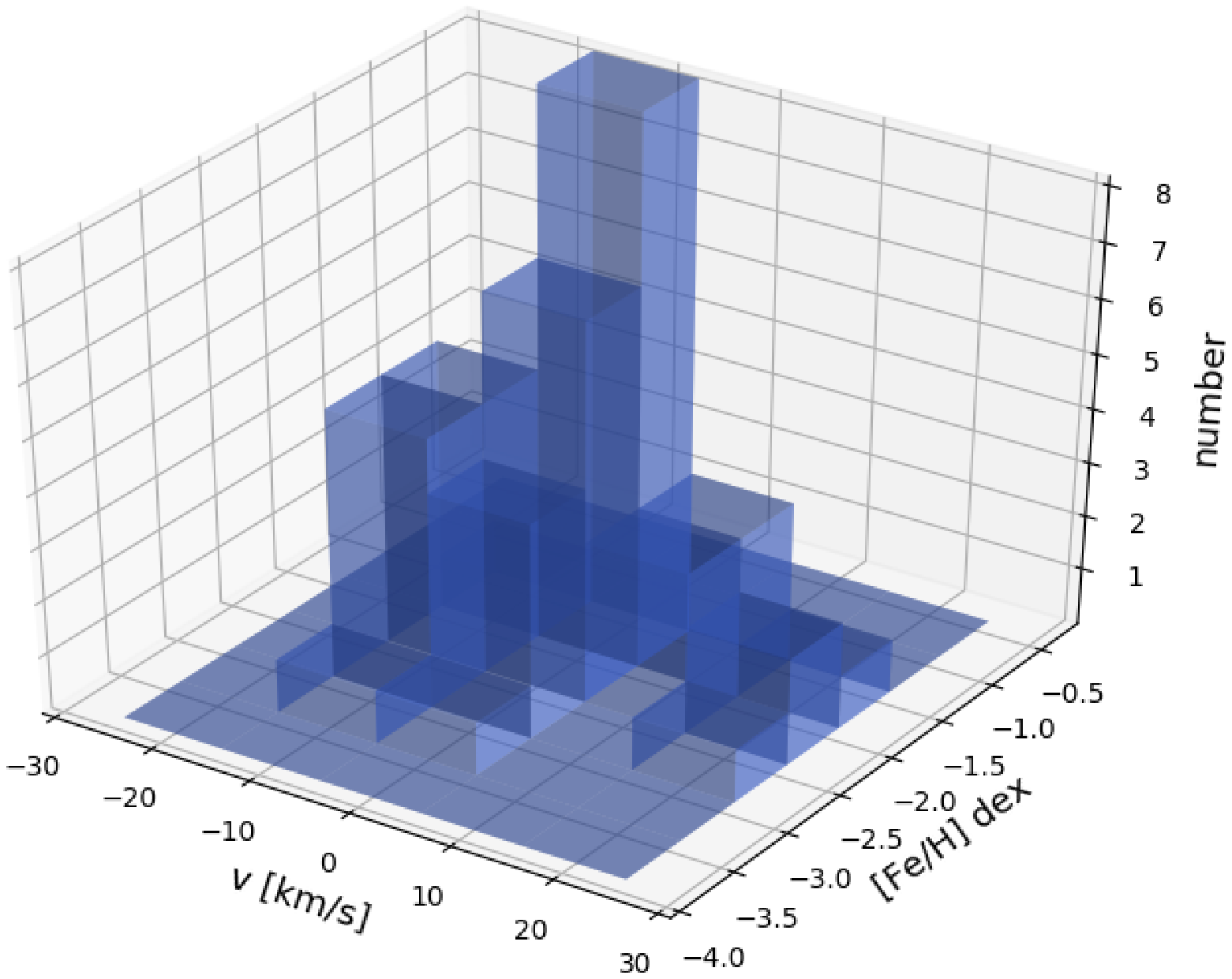}}}\\
(c){{\includegraphics[width = 7cm]{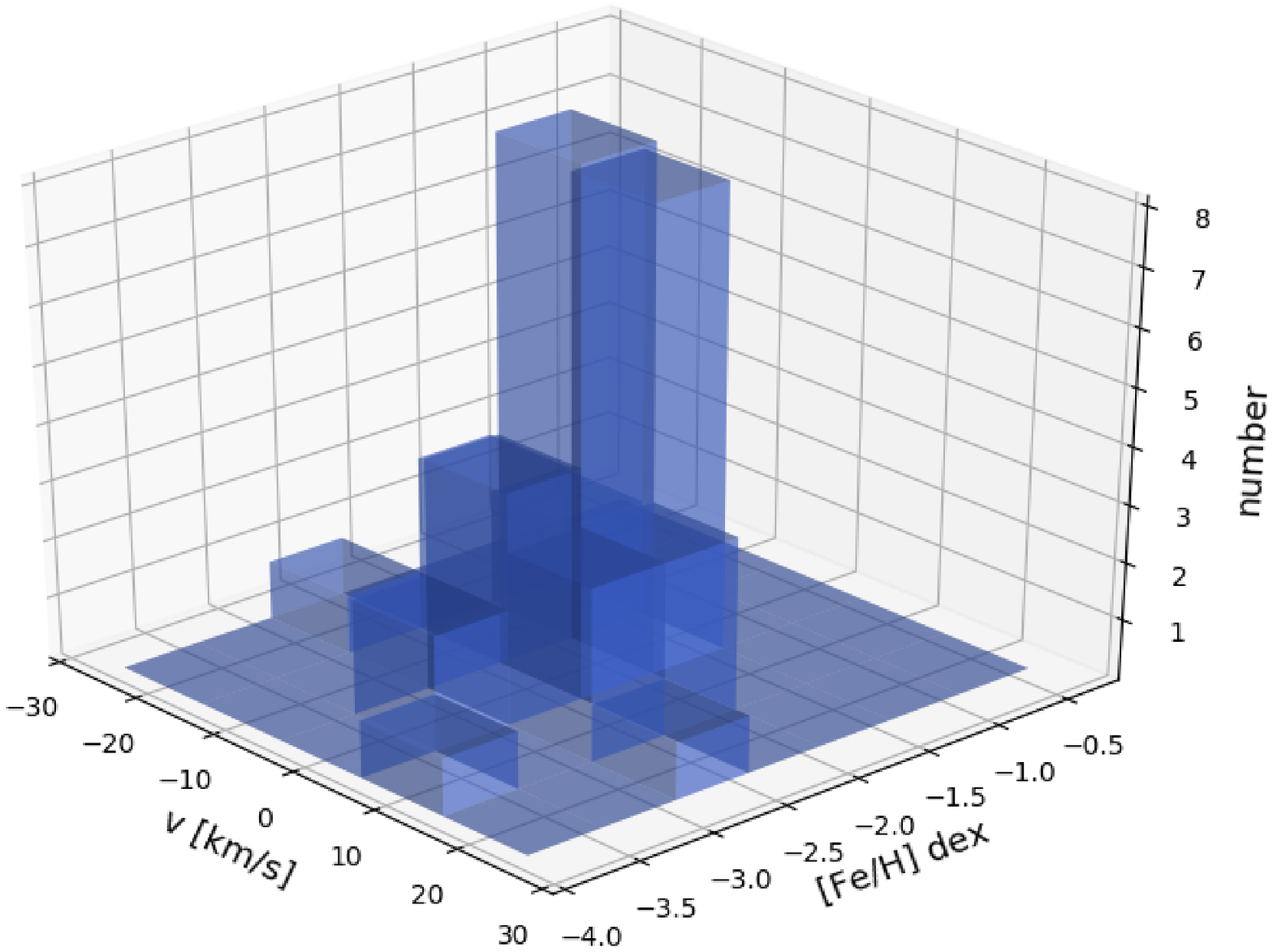}}}
(d){{\includegraphics[width = 7cm]{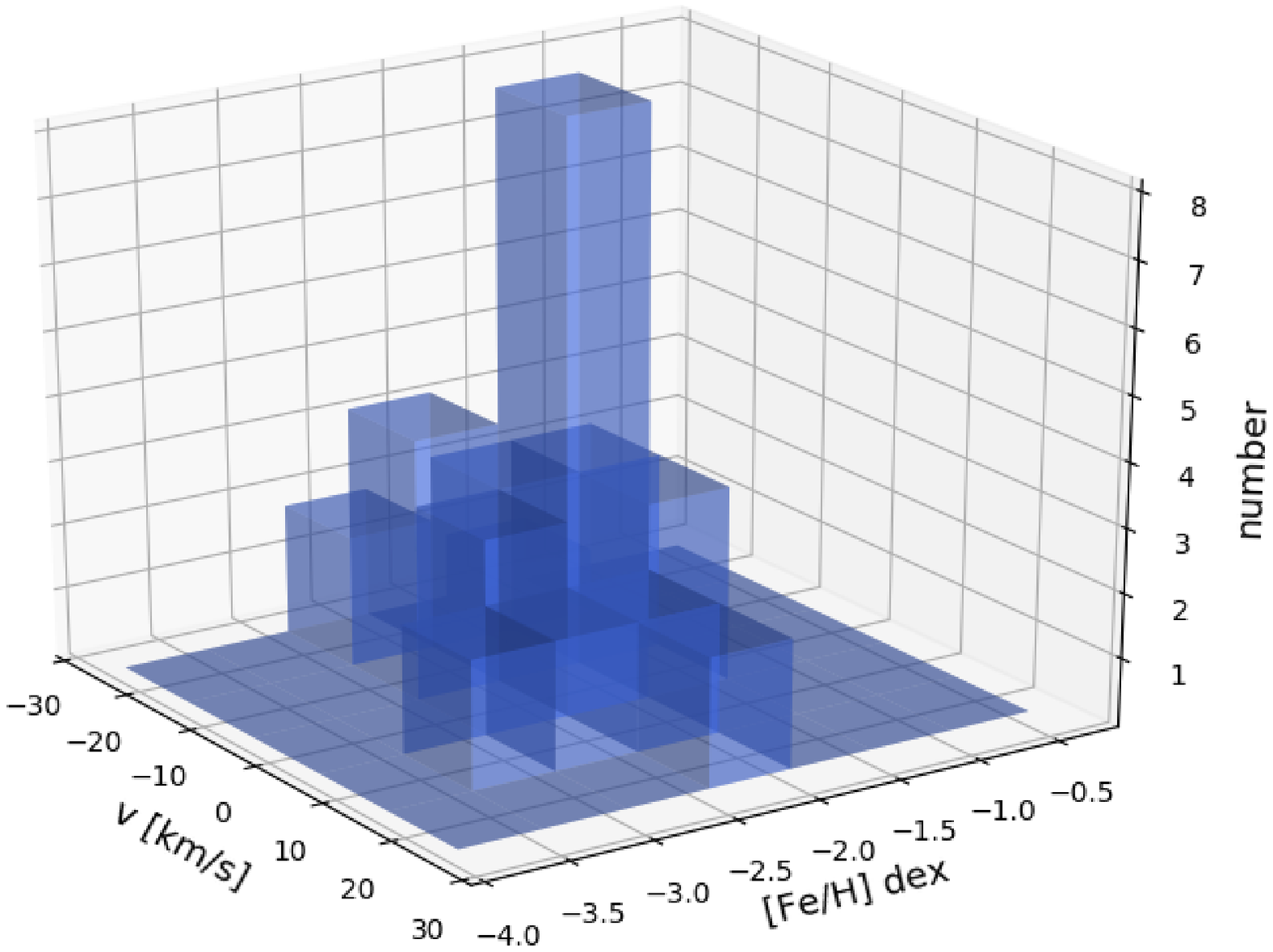}}}
\caption{Brute-force method: 2D Histograms of metallicity and radial velocity
per elliptical annulus bin. As an example, using 
\citeauthor{battaglia:11}'s \citeyear{battaglia:11} data, 
we show the 2D histograms for four elliptical-radial bins for the 
Sextans dwarf; (a) [0-10]~arcmin, (b)~(10-20]~arcmin, (c)~(20,30]~arcmin, 
and (d) (30,40]~arcmin. The typical size of the uncertainty in metallicity is $-0.2$~dex,
and the typical size of the uncertainty in the radial velocity is $2.1$~km~s$^{-1}$.}
\label{Fig:2d_histograms}
\end{figure*}

\subsection{Leo I}
\label{sec:leoI}


The Leo I dSph is one of the most remote dSphs associated with the
MW (e.g. \citeauthor{grebel:03} \citeyear{grebel:03}). It has a Galactocentric 
distance of $254$~kpc \citep{bellazzini:04}, a position angle $PA=78$\textdegree, a core radius 
$r_{c}=3.6\pm0.1$~arcmin, a tidal radius $r_{t}=13.5\pm0.3$~arcmin, an 
ellipticity $e=0.31$ \citep{munoz:18}, and a systemic velocity $v_{sys}=284.2$~km~s$^{-1}$
\citep{koch:06}.

Because of Leo~I's high radial velocity and its large Galactocentric distance, 
\cite{koch_leoI:07} argue that Leo~I might be an isolated system, which is currently 
not affected by Galactic tides. 
However, \cite{boylan-kolchin:13} argue that it is very unlikely that Leo~I is not 
bound to the MW galaxy, under the premise that Galactic satellites are associated 
with DM subhalos. They used high resolution numerical simulations of a 
MW like DM halo and found that 99.9\% of the subhalos in the simulations 
are bound to their host halo.

On the other hand, if Leo~I passed very close to the Galactic center (around $1$~Gyr 
ago), then the observed kinematics and population segregation 
in Leo~I, along with its distorted structural parameters, its age, its last prominent 
burst of star formation, and its large radial velocity relative to the Galactic Center, 
can be explained \citep{mateo:08}. Because Leo~I presents no tidal arms, the 
latter scenario could result from the interaction with a third body, which placed
Leo~I into its present highly-elliptical orbit.

\cite{sohn:13} studied the detailed orbital history of Leo~I. They found that Leo~I 
entered the MW virial radius $2.3$~Gyr ago, and that 
(confirming \citeauthor{mateo:08} \citeyear{mateo:08}) it had a pericentic approach 
(at a Galactocentric distance $D_{GC}=91$~kpc) around one Gyr ago.    

However, \cite{koch:12b} measured the proper motion of Leo~I and found that Leo~I might 
not be bound to the MW. Furthermore, they say that it is likely that Leo~I was formed 
and evolved in isolation, and it is now approaching its first encounter with the Galactic 
halo.

\cite{gaia:18} estimated that Leo~I has a period greater than $5$~Gyr, and they 
predict that its last pericentric passage (at a distance of $\sim100$~kpc) took place
around $1$~Gyr ago.
The Leo~I dSph has been used to derive a limit on the mass of the MW DM halo. 
If Leo~I is indeed bound to the MW, it sets constrains on the MW mass 
\citep{boylan-kolchin:13}. \cite{gaia:18} set a lower limit on the mass of
the MW DM halo of $9.1^{+6.2}_{-2.6} \times10^{11}$~M$_{\odot}$, based on 
the assumption that Leo~I is indeed bound to the MW.
Recentely, \cite{fritz:18} found that Leo~I is ``back-splashing'' if one considers the 
(preferred) heavy MW DM halo model ($M_{MW}=1.6\times10^{12}$~M$_{\odot}$).

Interestingly, stellar substructure has been detected in Leo~I. \cite{mateo:08} reported 
six stars uniformly distributed and kinematically distinct from the main Leo~I stellar 
component. They claim that these stars might represent a tidal feature,  but they warn 
that the statistics are too poor and that further members of this kinematic structure 
would need to be identified in order to conclude whether the substructure is real or not.

\subsection{Leo II}
\label{sec:leoII}
The Leo II dSph is located at a Galactocentric distance of $233$~kpc \citep{bellazzini:05}.
It has a position angle $PA=12$\textdegree, a core radius $r_{c}=2.25\pm0.1$~arcmin,
a tidal radius $r_{t}=9.82\pm0.4$~arcmin, an ellipticity $e=0.07$ \citep{munoz:18}, 
and a systemic velocity in the heliocentric system $v_{sys}=223.9$~km~s$^{-1}$ \citep{koch:06}.

Leo~II is believed not to be experiencing strong Galactic tides, and its proper
motion indicates that it might not even be a bound satellite to the MW \citep{lepine:11}. 
\cite{piatek:16} measured the proper motion of Leo~II. They found that the motions
they measured support the idea that Leo~II fell into the MW DM halo as a part of
a group. On the other hand, \cite{gaia:18} concluded that the infall of Leo~II as
a group is unlikely.

\cite{vogt:95} calculated a $M/L\approx7$, which indicates that Leo~II must be embedded in 
a DM halo, but it is not an extreme case.
\cite{koch_leoII:07} obtained a large data set of radial velocity measurements 
out to Leo~II's limiting radius. They found (depending on the total luminosity adopted) a 
$M/L=25-50$. They concluded that this $M/L$ ratio together with the flatness of its 
dispersion profile indicate that Leo~II is a DM-dominated system. 

\cite{komiyama:07} reported the existence of a small stellar substructure beyond Leo~II's
tidal radius. The substructure's luminosity compares to that of a globular cluster ($M_{V}<-2.8$).
The substructure might be a disrupted globular cluster or a group of stars that were stripped 
away from the galaxy.

\section{Methods}
\label{sec:methods}
\subsection{``Brute-force'' method}
\label{sec:methods:brutus}

\begin{figure*}
    \centering
    (a){{\includegraphics[width=8.2cm]{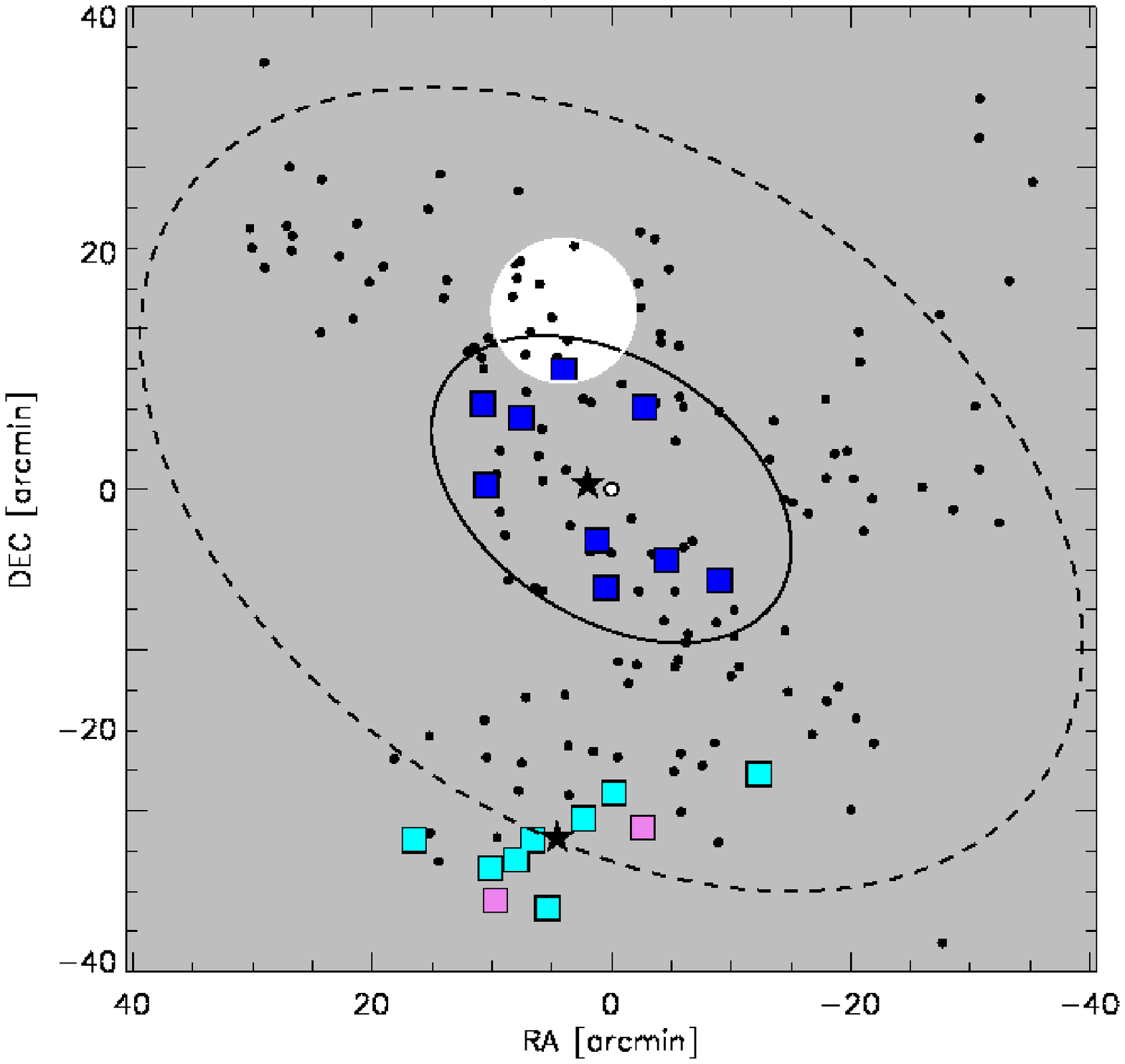} }}
    \qquad
    (b){{\includegraphics[width=8.2cm]{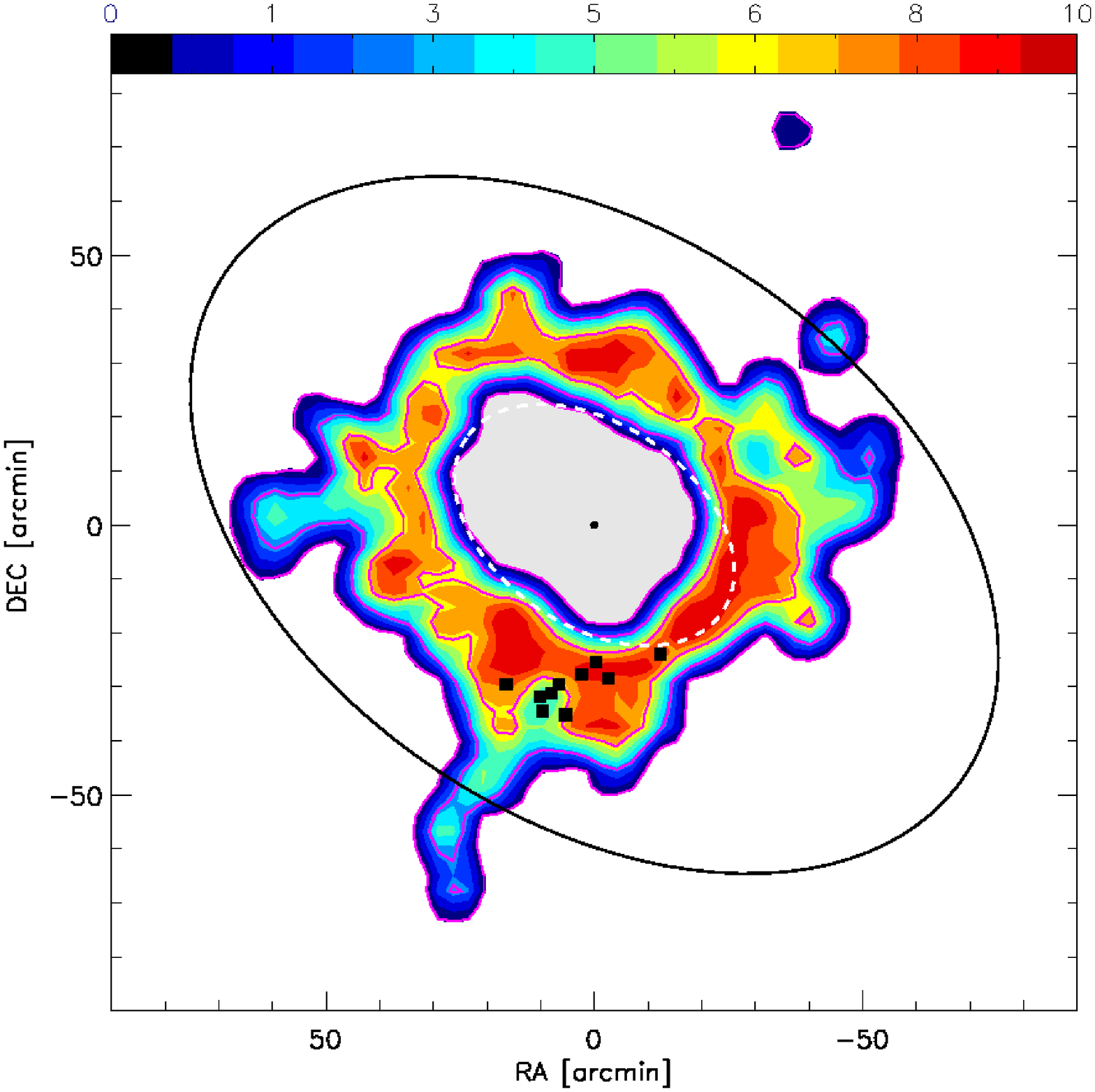} }}
    \caption{(a) The Sextans dSph members reported by \cite{battaglia:11} are shown 
    as filled-black circles. The white small circle (in the center) represents 
    the center of Sextans reported by \cite{mateo:98}. The stellar substructure previously found by 
    \cite{battaglia:11} (called \textsc{sxt1} in the text), is shown with 
    filled blue squares. The new stellar substructure, reported in this work 
    (\textsc{sxt2}) is shown with filled light blue squares (and two extra violet 
    squares, see text in Section~\ref{sec:sextans}). The solid black stars, in both 
    substructures (\textsc{sxt1} and \textsc{sxt2}) correspond to the centroid 
    of each stellar substructure. The solid ellipse correspond to the core radius 
    reported by \cite{irwin:95}. The dotted ellipse is placed at the center of 
    \textsc{sxt2} substructure, which corresponds to $43.5$~arcmin ($1$~kpc) from 
    the center of Sextans. The region marked in white shows the kinematically colder 
    region detected by \cite{walker:06}.\\
    (b) In this panel we show contours of star counts similar to 
    \citeauthor{roderick:16}'s (\citeyear{roderick:16}) Figure~13. We over plotted 
    their core ($26.8$~arcmin, see white dotted ellipse) and 
    their tidal radius ($83.2$~arcmin, see black ellipse), which they found 
    by fitting a King model to the radial distribution of Sextans. The substructure named \textsc{sxt2} is 
    shown as black filled squares.}
    \label{fig:sextans}
\end{figure*}

We search for stellar substructures in the four selected dSph galaxies. 
For this purpose, we first constructed 2-dimensional histograms on a grid of
metallicity vs. velocity (see Figure~\ref{Fig:2d_histograms}) for
every elliptical annuli of constant ellipticity and position PA in the 
selected galaxy. 


We adopt an initial metallicity $met_{i}$ (the minimum metallicity 
value of the studied data) and a final metallicity $met_{f}$ 
(the maximum metallicity in the data), with an interval in metallicity
$\Delta met$ and an interval in velocity of $\Delta v$, see Figure~\ref{Fig:2d_histograms}.


Subsequently, with both metallicity ($met$) and velocity ($v$) fixed, 
we count the stars in our data, that satisfy being in the interval
\begin{equation}
\label{eq:1}
 met  < met_{*} \leqslant met + \Delta met \mbox{ ,}
\end{equation}
where $met_{*}$ is the metallicity of the star that is being analyzed. We repeat the
procedure with the velocity, such that
\begin{equation}
\label{eq:2}
 v  < v_{*} \leqslant v + \Delta v \mbox{ ,}
\end{equation}
where $v_{*}$ corresponds to the velocity of the star that is being 
analyzed. With the resulting number of stars that satisfy the conditions \ref{eq:1} and 
\ref{eq:2} (per elliptical annulus width), we build histograms of the number 
of stars to find peaks in the counts of stars that could be interpreted as stellar 
substructures.

The peaks in the counts of stars that are interesting to us are the ones with 
low metallicities, and that are dynamically cold, meaning that the velocity dispersion 
of such group of stars should be significantly lower that that of the complete stellar sample 
(per elliptical annulus bin).

In order to declare whether a peak in counts is significant or not, we 
realize Monte Carlo tests of normal distributions for both metallicity and velocity 
(in the same elliptical annulus bin where the peak was found) and find the average and 
standard deviation ($\sigma$) of these distributions. If the number of counts in the 
peak (which is found in a defined metallicity-velocity cell) is $\geq~1.96\sigma$, or  
in other words, the probability of finding a peak $>1.96\sigma$ is $\leq5\%$ 
we take such a count peak to be significant.

\subsection{Minimum spanning tree method}
\label{sec:methods:tree}

Another method to investigate whether stars of a certain metallicity range are in some 
way kinematically clumped (i.e., spatially more concentrated) we make use of a minimum 
spanning tree (MST, e.g., \citeauthor{schmeja:11} \citeyear{schmeja:11}). 
The MST is the unique set of straight lines (``edges'') connecting a given set of points 
(``vortices'') without closed loops, such that the sum of the edge lengths is minimum. 
This construct from graph theory has been widely applied in astronomy to cluster and 
structure analysis, from the large-scale distribution of galaxies to the internal 
structure of star clusters \citep[and references therein] {schmeja:11}. We apply an 
approach similar to the one introduced by \cite{allison:09} to identify and quantify mass
segregation in star clusters. We construct the MST for the stars of a given metallicity 
range and determine the mean edge length $\gamma_{\rm mp}$ for those stars. We use the 
geometric mean rather than the arithmetic mean in order to minimize the influence of 
outliers \citep{olczak:11}. Then we construct the MST of the same number of randomly 
selected stars from the entire sample and determine the mean edge length 
$\gamma_{\rm rand}$. This is done 200 times in order to obtain the mean value 
$\langle \gamma_{\rm rand} \rangle$.
The ratio 
\begin{equation}
\label{eq:ratio}
R = \frac{\langle \gamma_{\rm rand} \rangle}{\gamma_{\rm mp}} \mbox{ ,}
\end{equation}
is a measure for the concentration of the stars of the sub-sample relative to the entire 
stellar population. A value of $R \approx 1$ implies that both samples are distributed in a 
similar manner, while $R \gg 1$ indicates that the sub-sample is more concentrated than the 
sample as a whole.

\section{Results}
\label{sec:results}
\subsection{The case of Sextans}
\label{sec:results_sextans}
For the case of Sextans, we used \citeauthor{battaglia:11}'s (\citeyear{battaglia:11}) data. They obtained VLT/FLAMES 
intermediate-resolution 
spectroscopic observations in the near-infrared CaII triplet (CaT) region for $1036$ 
distinct targets along the line-of-sight to Sextans. The magnitudes and colors of 
those targets are broadly consistent with red giant branch (RGB) stars. From that sample they obtained 
$789$ stars with S/N and error in velocity that produce reliable line-of-sight velocities 
and CaT equivalent widths. A subset of $174$ stars from those are RGB stars that are probable members 
with line-of-sight velocities accurate to $\pm 2$~km/s and CaT [Fe/H] measurements accurate 
to $\pm 0.2$~dex. The majority of the Galactic contaminants were eliminated from \citeauthor{battaglia:11}'s
sample using a $3\sigma$ kinematic cut. Finally, in order to refine their membership criteria, 
they used the Mg~I line at $8806.8$ \AA{} as an empirical indicator of stellar surface 
gravity, so that the probable members (RGB stars) of Sextans and the Galactic contaminants could be
distinguished.

Our first target of study is the stellar clump found in Sextans by \cite{battaglia:11}. 
We will refer to this clump of stars as \textsc{sxt1}. With our {\em brute-force} method (see 
Section \ref{sec:methods:brutus}), we should be able to reproduce 
previous findings. Using the \cite{battaglia:11} data, we follow the method of
Section~\ref{sec:methods:brutus}. We set  $met_{i}=-3$~dex, $met_{f}=-0.28$~dex 
and $\Delta met=0.5$~dex. For the velocities, we set $v_{i}=-20$~km~s$^{-1}$,  
$v_{f}=20$~km~s$^{-1}$, and a $\Delta v=10$~km~s$^{-1}$.


Then we construct histograms of star counts per elliptical radius. These 
histograms help us to identify the accumulation of stars in a particular radius bin. 
The elliptical radial bins in the case of Sextans have a size of 10~arcmin. 
As expected, we find \textsc{sxt1}.

The metallicity of this group of stars (\textsc{sxt1}) is in the range 
$-2.84 \leq met \leq -2.43$~dex, with an average metallicity [Fe/H]$=-2.64$, and the 
velocity is in the range $-2 \leq v \leq -12$~km~s$^{-1}$. The fact that we are able to recover
the \textsc{sxt1} clump is an indicator that our method is working well. 

In Figure~\ref{fig:sextans}a we plot Sextans' member stars (black circles) taken from the 
\cite{battaglia:11} data set. The center of the Sextans dSph \citep{mateo:98} is shown as a 
white circle. The \textsc{sxt1} clump stars are shown as filled blue squares, and the 
solid black star shows the centroid of \textsc{sxt1}. 

We did not only recover the \textsc{sxt1} clump reported by \cite{battaglia:11}, 
but we also found a second old cold-stellar substructure.
We will refer to this new substructure as \textsc{sxt2}. The \textsc{sxt2} stellar 
substructure consists of eight stars with metallicities ranging from $-2.99$ to $\leq -2.63$.
The average metallicity is [Fe/H]= -2.78, and the velocities range from 
$3.7 \leq v \leq 10.08$~km~s$^{-1}$. We computed the centroid (relative to the center of Sextans, 
see Table \ref{table:parameters}) of \textsc{sxt2} to be located at ($4.537$,$-29.265$)~arcmin 
(relative to the center of Sextans).


In Figure~\ref{fig:sextans}a, we show the stars of the new clump 
(\textsc{sxt2}) as filled cyan squares, and the respective centroid plotted
as a solid black star. It has to be noted that if one relaxes the velocity constraint 
and lets the velocity cover $3.7 \leq v \lesssim 16$~km~s$^{-1}$, we can add two 
new metal-poor stars. Therefore, we end with a 10-star substructure with an average 
metallicity of $-2.76$. If we include these two stars, the centroid of the 
$10$-star substructure changes to ($4.334$, $-29.700$)~arcmin. The two extra 
stars are shown as two filled magenta squares in Figure~\ref{fig:sextans}a. 

The MST analysis also detects \textsc{sxt1} and \textsc{sxt2} and confirms them as 
highly concentrated clumps with $R = 3.28 \pm 0.20$ and $R = 3.53 \pm 0.31$, 
respectively (see Equation~\ref{eq:ratio}).

In order to demonstrate that random groupings of stars with matching 
metallicities and velocities are in fact very rare, we realize Monte Carlo tests 
of normal distributions for both metallicity and velocity (see 
Section~\ref{sec:methods:brutus}). We find that the probability of finding 
\textsc{sxt1} is $\sim1\%$ (i.e., $\sim2.5\sigma$).
We also find that the probability of finding \textsc{sxt2} is $\sim5\%$ 
(i.e., $2.1\sigma$).


The latter supports the suggestion that in fact the new clump 
\textsc{sxt2} reported in this work is likely physical.


\begin{figure*}
    \centering
    (a){{\includegraphics[width=8.2cm]{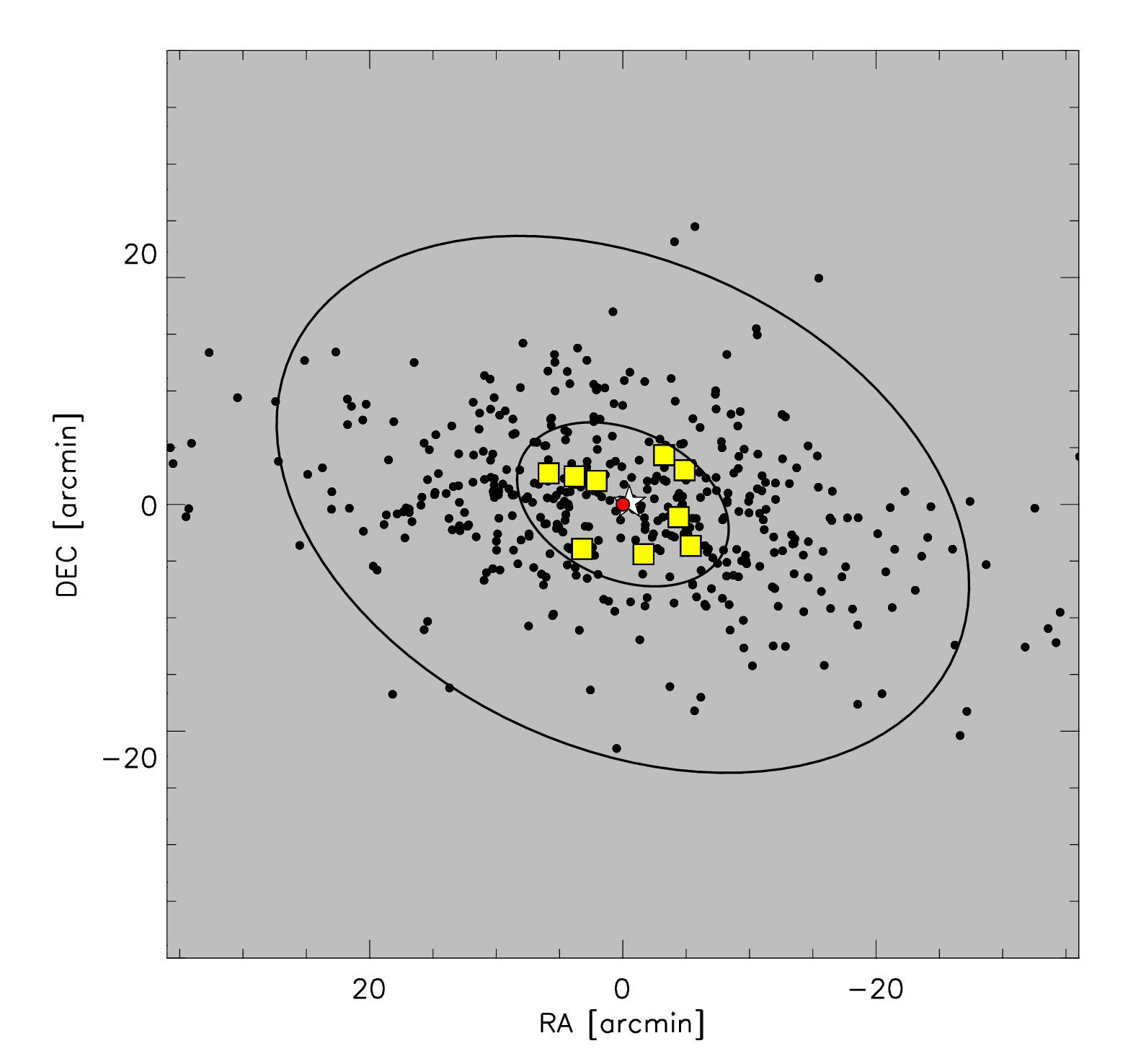} }}
    \qquad
    (b){{\includegraphics[width=7.6cm]{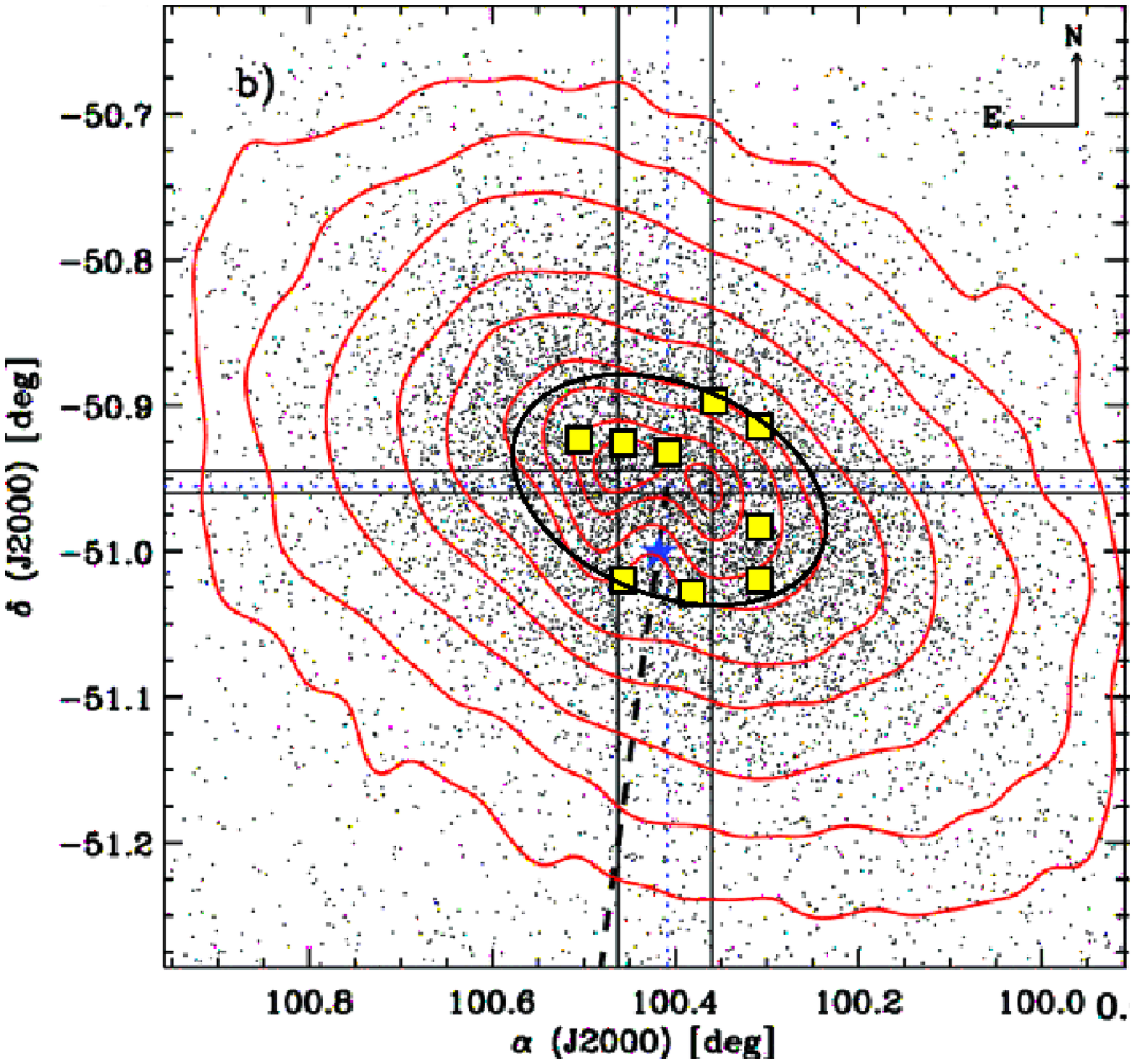} }}
    \caption{(a) In this panel we show the Carina dSph members reported by \cite{koch:08} 
as filled black circles. The center of Carina is represented with a filled red circle. 
The new stellar substructure found in this work (\textsc{car1}) is shown with filled yellow 
squares.The white star corresponds to the centroid of the stellar substructure (relative 
to Carina's center, see Table~\ref{table:parameters}).
The two solid black ellipses show the core radius at $8.8$~arcmin ($240$~pc) and the 
tidal radius at $28.8$~acrmin ($787$~pc).\\
    (b) \citeauthor{fabrizio:11}'s (\citeyear{fabrizio:11}) Figure~13: sky distribution 
    of the stars from \citeauthor{fabrizio:11}'s (\citeyear{fabrizio:11}) 
photometric catalog. Isodensity contours are shown ranging from $5$ to $95$\% of 
the total number of stars. 
The blue star shows the position of the star HD48652. The solid (vertical and horizontal) 
black lines mark the two secondary peaks in RV. The nine stars making up \textsc{car1} are 
overplotted as filled yellow squares. The black solid ellipse is located at the distance of the
core radius ($8.8$~arcsec).}
    \label{fig:carina}
\end{figure*}
%

Interestingly, \cite{roderick:16} reported clear evidence of a stellar substructure distributed 
evenly about the center of Sextans. The substructure extends up to a distance of $2$~kpc. 
In order to compare our findings to \cite{roderick:16}, we built contours of star counts 
using \citeauthor{roderick:16}'s (\citeyear{roderick:16}) substructure data, shown in Figure~\ref{fig:sextans}b.
We over-plotted their core (white dashed-line) and tidal radius (black line) 
($26.8$~arcmin and $83.2$~arcmin, respectively). It is very encouraging to observe that 
our \textsc{sxt2} substructure lies in a dense substructure region, giving a hint that \textsc{sxt2}
might be a member of the annular substructure within Sextans.

\cite{battaglia:11} transformed the heliocentric line-of-sight velocities into line-of-sight 
velocities in a frame at rest with respect to the Galactic center, $v_{GSR}$ (where GSR 
stands for Galactic standard of rest, $v_{sys,GSR}=78.4 \pm0.6$~km~s$^{-1}$). 
They reported a velocity dispersion for the six innermost
stars making up their clump of $1.4\pm1.2$~km~s$^{-1}$, and an average GSR velocity of 
$72.5\pm1.3$~km~s$^{-1}$. They also computed an average metallicity for their clump of [Fe/H]= -2.6.

We follow \cite{battaglia:11} and compute a velocity dispersion and a mean GSR velocity for 
eight stars of \textsc{sxt2} (see the cyan squares in Figure~\ref{fig:sextans}a) of $2.2$~km~s$^{-1}$ and 
$87.1$~km~s$^{-1}$, respectively. 

Since the velocity dispersion could be a strong function of the galactocentric radius, 
we compute the velocity dispersion of all the stars in \citeauthor{battaglia:11}'s 
(\citeyear{battaglia:11}) data belonging to the same elliptical annulus as \textsc{sxt2}. 
We found that the velocity dispersion of the elliptical annulus associated with \textsc{sxt2} 
is 9~km~s$^{-1}$.

The velocity dispersion of \textsc{sxt2} is very cold (compared to the velocity dispersion 
of all the stars belonging to the same elliptical annulus) and its average metallicity ([Fe/H]= -2.78) is
even lower than that of \textsc{sxt1}. The low metallicity of \textsc{sxt2}, together with its cold 
kinematics, suggests that the stars belonged to a stellar (globular) cluster. 
If we add the two extra stars shown as magenta squares in Figure~\ref{fig:sextans}a (where we 
have relaxed the velocity condition, see text above), we obtain a velocity dispersion of 
$4.2$~km~s$^{-1}$ and an average metallicity of [Fe/H]=-2.76. The velocity dispersion of 
the 10-star clump is a bit higher than the one computed with eight stars, but it remains 
colder than the velocity dispersion of \textsc{sxt2}'s associated elliptical annulus  
($9$~km~s$^{-1}$).

As done by \cite{battaglia:11}, if the $174$ stars in the Sextans sample are representative, 
then \textsc{sxt2} would account for $4.6\%$ (or $\sim 6\%$ if one includes the extra two stars) 
of Sextans' stellar population. If we take the value for the luminosity of Sextans of $4.37\times10^{5}$~L$_{\odot}$ 
\citep{irwin:95,lokas:09} and a typical value for the mass-to-light ratio for globular 
clusters, $M/L\sim2$ \citep{mclaughlin:05}; then zero order estimates for the mass and luminosity of 
\textsc{sxt2} are \mbox{$4\times10^{4}$~M$_{\odot}$} and $2\times10^{4}$~L$_{\odot}$ 
($5\times10^{4}$~M$_{\odot}$ and $2.5\times10^{4}$~L$_{\odot}$, if we include the two extra stars).   

The fact that the MST analysis confirmed that \textsc{sxt2} is concentrated with 
respect to the overall Sextans sample, along with its metal-poor, dynamically cold 
nature, reasserts the idea that Sextans has not experienced a strong tidal encounter,
indicating that it could have a more circular orbit around the Galactic 
Center. The low metallicity of \textsc{sxt2} suggests that it might be very old 
(i.e., $>10$~Gyr) and its progenitor (as well as the progenitor of \textsc{sxt1}, 
\citeauthor{battaglia:11} \citeyear{battaglia:11}) would be most likely a globular cluster,
and such a globular cluster would then be among the most metal-poor globular clusters known.

The detection of cold old stellar substructures in dwarf spheroidal galaxies has 
been helpful to shed some light on the core/cusp DM problem at galactic scales. 
For example, a cored DM halo profile is preferred when explaining the existence of
the stellar substructures in UMi and Sextans \citep{kleyna:03,lora:09,lora:13}. 
The stellar substructure \textsc{sxt2} gives a new test to corroborate if (in 
particular) Sextans is embedded in a cored DM halo.\\

\subsection{The case of Carina}
\label{sec:results_carina}
%
For the case of Carina, we used \citeauthor{koch:06}'s (\citeyear{koch:06}) data.
Their targets in Carina were chosen from photometry and astrometry obtained by 
the ESO Imaging Survey \citep{nonino:99}. 
\cite{koch:06} selected their targets to cover magnitudes from the tip of the RGB 
down to $3$~mag below the RGB tip ($20.3$~mag in apparent V-band magnitude).
They chose their RGB magnitude range such that even for the faintest stars they
would be able to achieve high signal-to-noise ratios.
They obtained VLT/FLAMES spectroscopic observations in the near-infrared 
CaT region, using the GIRAFFE spectrograph  with both MEDUSA fiber slits 
in ``low-resolution'' mode. Their observed fields covered most of 
Carina's area, and extended even beyond its nominal tidal radius.
The membership probabilities were calculated using an error-weighted
maximum-likelihood fit of a Gaussian velocity distribution (in the
$150-300$~km~s$^{-1}$ range), and then rejecting $3\sigma$ outliers.
\cite{koch:06} determined the metallicities of the RGB stars sample through the 
equivalent widths-CaT method.

We apply the same methods described in Section~\ref{sec:methods} to the 
Carina dSph. 
We set the metallicity limits $met_{i}=-3$~dex, $met_{f}=-0.3$~dex and 
$\Delta met=0.5$~dex. For the velocities we set $v_{i}=-20$~km~s$^{-1}$ and 
$v_{f}=20$~km~s$^{-1}$ with a $\Delta v=10$~km~s$^{-1}$. 
We find a group of nine stars with metallicities in the range $-2.89 \leq met \leq -2.44$ 
and velocities $2.93 \leq v \leq 13.31$~km~s$^{-1}$, within elliptical 
annuli widths of 5~arcmin. We refer to this group of  stars as \textsc{car1}. 
We plot \textsc{car1} in Figure~\ref{fig:carina}a as filled yellow squares. 
\textsc{car1}'s centroid is shown with a white star. We notice that the centroid 
of \textsc{car1}($-0.504,0.153$~arcmin) is quite close to the center of Carina 
(just $\sim 14$~pc away), represented with a filled red circle.

Applying the MST analysis, \textsc{car1} is found as a prominent clump with 
a ratio $R = 3.51 \pm 0.30$ (see Equation~\ref{eq:ratio}).

We demonstrate that random groupings of stars with matching metallicities 
and velocities in Carina's data are very rare. We realize Monte Carlo tests 
of normal distributions for both metallicity and velocity (see 
Section~\ref{sec:methods:brutus}). We find that the probability of finding 
\textsc{car1} is $\sim4.5\%$ (i.e., $2.15\sigma$) supporting the suggestion 
that the substructure \textsc{CAR1} is physical.
The stellar substructure found in Carina could be an unbound object in the throes of 
destruction, which could have started out as a typical cluster with a radius of 
$\sim3-5$~pc, very similar to the stellar substructure \textsc{sxt1}.

For the nine stars group (\textsc{car1}) we computed a velocity dispersion of 
$4$~km~s$^{-1}$, a mean velocity of $230.2$~km~s$^{-1}$, and an average metallicity 
of [Fe/H]$\sim -2.6$. 

We compute the velocity dispersion of all the stars in \citeauthor{koch:06}'s 
(\citeyear{koch:06}) data belonging to the same annulus-bin as \textsc{car1}. 
We found that the velocity dispersion of the elliptical-annuli associated to \textsc{CAR1} 
is 7.2~km~s$^{-1}$.


If the $437$ star-sample is representative of Carina' s stellar component, 
then \textsc{car1} would account for $2\%$ of Carina's stellar population. If 
we adopt a mass-to-luminosity-ratio of $2$ (typical for globular clusters) and a 
luminosity of $3.5\times10^{5}$ \citep{lokas:09}, then we can roughly estimate that 
\textsc{car1}'s mass is $\sim1.4\times10^{4}$, very similar to the mass estimate of 
\textsc{sxt1} and \textsc{sxt2}.

As we mentioned before, \cite{fabrizio:11} found evidence of a secondary maximum 
in radial velocity at a distance $\sim200$~pc from Carina's center, which they interpreted
as reminiscent of a substructure with transition properties (i.e., a transition between 
a bulge-like and/or disk-like structure).
In Figure~\ref{fig:carina}b we took the sky distribution of the stars in 
\citeauthor{fabrizio:11}'s (\citeyear{fabrizio:11}) 
photometric catalog (their Figure~$13$), where they plot isodensity contour levels from
5\% to 95\%. We over plotted the \textsc{car1} clump of stars. All the (nine) stars lie inside 
the 45\% isodensity contour level, which coincides with the core radius of Carina (see the black 
ellipse in Figure~\ref{fig:carina}b), and it also coincides with the location where 
\cite{fabrizio:11} reported their substructure.
%

\subsection{The case of Leo~I}
\label{sec:results_LeoI}
 \begin{figure*}
    \centering
    (a){{\includegraphics[width=8.2cm]{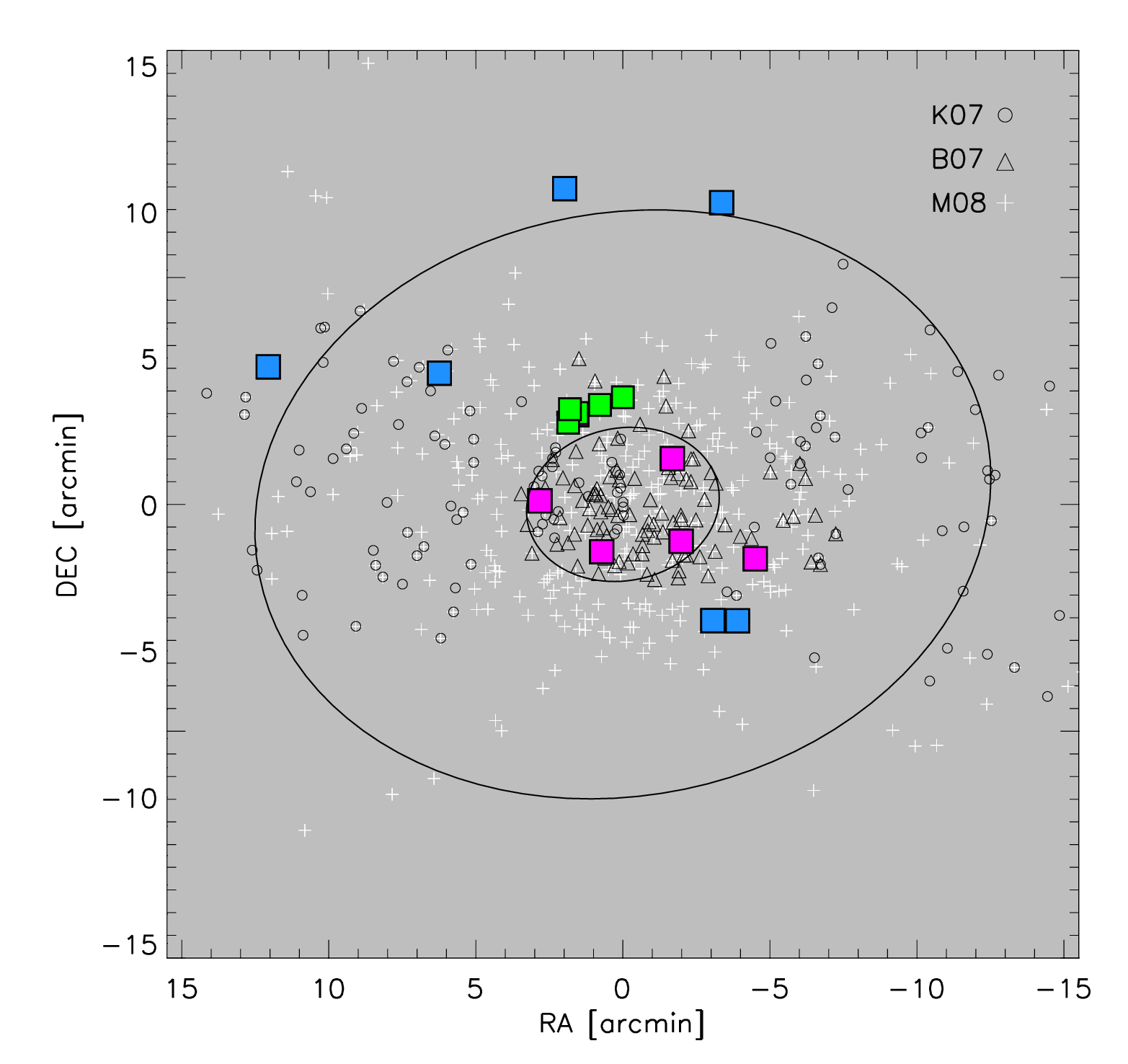} }}
    \qquad
    (b){{\includegraphics[width=8.2cm]{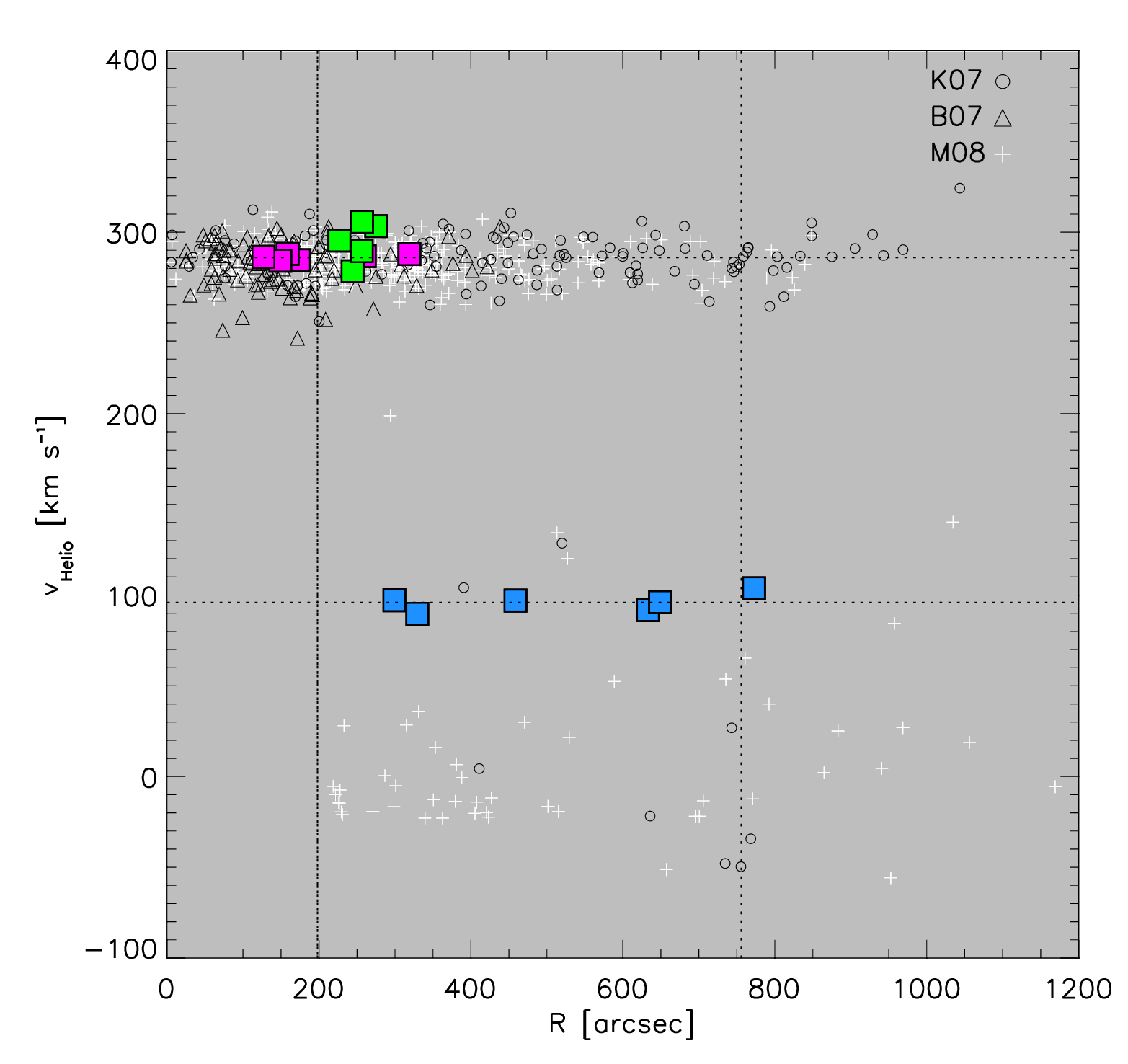} }}
     \caption{(a) Leo~I dSph members reported by \cite{koch_leoI:07} are shown as black open 
    circles. The red giant stars reported by \cite{bosler:07} are represented with black open triangles.
    The white plus symbols represent \cite{mateo:08}'s data. 
    Overplotted are the group of stars reported in this work \textsc{li1} (-filled-magenta squares);
    the group of stars reported by \cite{koch_leoI:07} (filled green squares),
    and the group of stars reported by \cite{mateo:08} (filled blue squares).
    The inner solid black ellipse corresponds to the core radius and 
    the outer solid black ellipse corresponds to the tidal radius \citep{munoz:18}. \\
    (b) Heliocentric velocities  as a function of the elliptical radius. The dotted
    vertical lines show the core ($3.3'$) and tidal ($12.6'$) radius. The horizontal 
    dotted line shows $v_{Helio}\approx96$~km~s$^{-1}$, the average velocity of \cite{mateo:08}'s 
    sample.} 
    \label{fig:leoI}
\end{figure*}
For the case of Leo~I, we used \citeauthor{koch_leoI:07}'s (\citeyear{koch_leoI:07}) 
(hereafter K07) data together with \citeauthor{bosler:07}'s (\citeyear{bosler:07}) 
(hereafter B07) data.

The targets in \cite{koch_leoI:07} were selected from photometry obtained with the
framework of the Cambridge Astronomical Survey Unit \citep{irwin:01} at the $2.5$~m 
Isaac Newton Telescope. They selected red giant stars in order to cover magnitudes 
from the tip of the RGB ($I\sim18$~mag) down to $1.6$~mag below the RGB tip ($I\lesssim19.6$~mag).
Their observations were carried out with the 
Gemini Multiobject Spectrograph (GMOS) and with the Deep Imaging Multi-Object Spectrograph 
(DEIMOS).
Their low-resolution spectra have a signal-to-noise of $\sim5$, sufficient to obtain 
accurate radial velocity measurements. They derived radial velocities from their final 
reduced spectra by cross-correlating the three strong Ca lines ($8498$, $8542$, and $8662$) 
with a synthetic template spectrum of the CaT region. They estimated the metallicities of 
their RGB stars sample, from the widely used method of relating the equivalent widths of the 
CaT to metallicity.

On the other hand, \cite{bosler:07} obtained low-dispersion spectra of red giants in Leo I 
(and Leo~II) using the Keck I 10-m telescope and LRIS.
They obtained a mean S/N of $18$ (for Leo I stars). They verified the membership of their RGB 
stars by deriving their radial velocities. From their $121$ stars observed in Leo~I, 
a number of $90$ have heliocentric velocities within $3\sigma$ of the average velocity, 
and thus were selected as members of Leo~I.
They estimated the metallicities of their 
sample relating the equivalent widths of the CaT (as done by \citeauthor{koch_leoI:07} 
\citeyear{koch_leoI:07}).

It has to be noted that the K07 and B07 data sets do not overlap.

We took both data sets, and applied the brute force method 
(Section~\ref{sec:methods:brutus}) to scrutinize the possible existence of 
stellar substructure. 
We set the ranges for the metallicity $met_{i}=-2.7$~dex, $met_{f}=-0.3$~dex 
and $\Delta met=0.5$~dex. For the velocities we set $v_{i}=-20$~km~s$^{-1}$ and 
$v_{f}=20$~km~s$^{-1}$ with $\Delta v=10$~km~s$^{-1}$. 

We only found one group of stars that was relevant, using elliptical annuli 
widths of 4~arcmin. This group of stars consists of six stars (hereafter, 
\textsc{li1}). The metallicity of \textsc{li1} is in the range 
$-1.78 \leq met \leq -1.52$, with an average metallicity [Fe/H]$= -1.63$. 
The velocities are in the range $288 \leq v \leq 284.3$~km~s$^{-1}$, the average 
velocity is $286$~km~s$^{-1}$, and the velocity dispersion is $1.6$~km~s$^{-1}$.

We computed the velocity dispersion of all the stars in the K07 and 
B07 data belonging to the same elliptical annulus as \textsc{LI1}. We found that the 
velocity dispersion of the elliptical annulus associated with \textsc{LI1} 
is 13~km~s$^{-1}$.

We show the stars belonging to \textsc{li1} as filled magenta squares in 
Figure~\ref{fig:leoI}. The open black circles are the red giants reported 
by K07 and the open black triangles correspond to B07's stars.

The MST method only finds few rather weak concentrations with $R \lesssim 1.5$ and 
a low number of members, therefore it did not help us to corroborate if \textsc{li1} 
is a real substructure.

K07 detected a minor significant rise in the radial dispersion
profile, around $~220$~pc from the center of Leo~I (corresponding to the core radius), 
which they not find to be associated with any real localized kinematical substructure (we will 
refer to this group of stars as \textsc{k1}). We plot this group of five stars  
as filled green squares in  Figure~\ref{fig:leoI}. We observe that the clump reported in 
this work and \textsc{k1} only have one star in common.
Adding to K07's and B07's data, additional data of Leo~I star members, 
\cite{mateo:08} (hereafter M08) presented 
kinematic results of stars in the Leo~I dSph (see white plus symbols in 
Figure~\ref{fig:leoI}). They found a group of six stars (see 
filled blue squares in Figure~\ref{fig:leoI}) with velocities in a 
narrow range, with a dispersion of $\sim5$~km~s$^{-1}$ and a mean velocity of
$\sim 96$~km~s$^{-1}$ (see horizontal dotted line in Figure~\ref{fig:leoI}b). 
They argue that this group of stars might be a kinematically cold group, but they 
warn that the statistics are poor. They found this
group of stars by plotting the heliocentric velocity as a function
of the elliptical radius (see our Figure~\ref{fig:leoI}b and Figure~9 of 
M08). For comparison, we over plotted the clump of stars reported
in this work, \textsc{li1} (magenta filled squares); and the group of five stars 
reported by K07 (filled green squares).

We perform a statistical study to show whether random groupings 
of stars in K07 and B07 data are rare or not. We realize Monte Carlo 
tests of normal distributions for both metallicity and velocity (see 
Section~\ref{sec:methods:brutus}). We find that the probability of finding 
\textsc{li1} is $\sim75\%$ (i.e., $0.33\sigma$). Such a finding reflects
the fact that, in the case of Leo~I, we are dealing with low number statistics, 
and thus \textsc{li1} is not statistically significant.

The velocity dispersion of \textsc{li1} is a factor of $\sim3$ smaller 
($\sigma_{v,LI1} = 1.6$~km~s$^{-1}$) than the dispersion found in M08's group 
of stars. This, together with the low average metallicity found in \textsc{li1},
indicates that \textsc{li1} (if real) might be a dynamically cold stellar substructure. 

Five of the six stars in \textsc{li1} are contained in the B08 data. Then, if we suppose that 
the B07's data are representative of the Leo~I main stellar population, \textsc{li1} would 
account for $\sim5\%$ of it. Depending on the luminosity adopted \citep{mcconnachie:12}, 
a zero order mass estimate for \textsc{li1} would be $M_{LI1}\approx(3.4-5)\times10^{5}$~M$_{\odot}$.
The calculated mass for \textsc{li1} is higher than for the stellar substructures
in Sextans and Carina (see Section~\ref{sec:results_sextans} and 
\ref{sec:results_carina}, respectively), but it is comparable to the mass of the most
massive globular cluster in the Fornax dSph and Sagittarius dSph, $\sim(2-4.5)\times10^{5}$~M$_{\odot}$
\citep{mackey:03}.
%
\begin{figure*}	
\epsscale{0.6}
\plotone{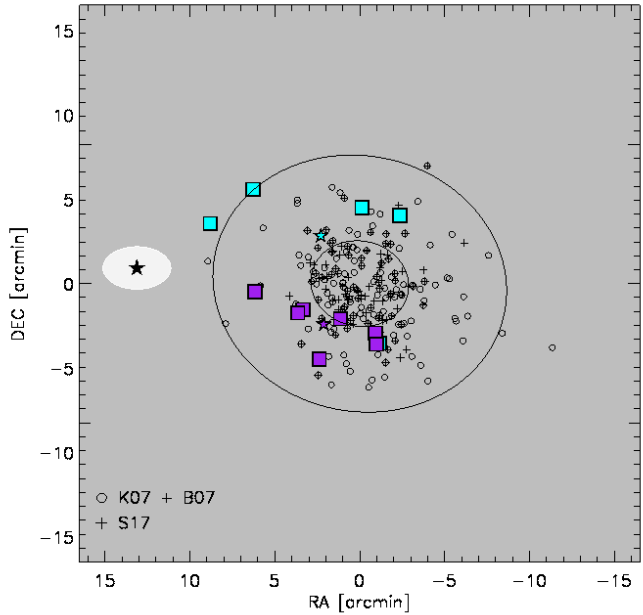}
\caption{We show the sky position of K207 and B07's Leo~II data as open circles, and S17 
data as plus symbols. The two substructures found in the S17 data are shown as 
filled cyan squares, and filled purple squares. The (cyan/purple) stars show 
the position of the centroid of each of the substructures. The two sold ellipses 
show the core (2.9') and tidal (8.7') radius. The black star shows the position 
of the center of the substructure reported by \cite{komiyama:07}, the white 
region shows the extent of the substructure ($4\times2.5$~arcmin$^{2}$).}
\label{fig:LeoII_ALL}
\end{figure*}
\subsection{The case of Leo~II}
\label{sec:LeoII}
For the case of Leo~II, we used \citeauthor{koch_leoII:07}'s (\citeyear{koch_leoII:07}) 
(hereafter K207) data, together with B07 data.

The targets in \cite{koch_leoII:07} were selected from photometry that was obtained by
the Cambridge Astronomical Survey Unit \citep{irwin:01} at the $2.5$~m Isaac Newton 
Telescope. The selection criteria of the targets involved colors and luminosities 
consistent with being members of Leo~II's RGB stars.
\cite{koch_leoII:07} obtained VLT/FLAMES spectroscopic observations in the near-infrared 
CaT region, using the GIRAFFE multifiber spectrograph in low-resolution mode, 
centered at the near-infrared CaT at $8550$~\AA{}.
To eliminate Galactic contaminants, \cite{koch_leoII:07} determined individual radial velocities 
by means of cross-correlation of the three CaT lines against synthetic Gaussian 
template spectra. The templates were synthesized adopting representative equivalent
widths of the CaT in RGB stars. 
The spectroscopic metallicity and age distributions were derived using the near-infrared
CaT calibration method \citep{koch:07c}.

On the other hand, B07 obtained CaT abundances and radial velocities for $74$ RGB stars in Leo~II, 
using the low-resolution spectrograph LRIS.
They obtained a mean S/N of $23$. They verified the membership of their RGB 
stars by deriving their radial velocities. From their $90$ stars observed in Leo~II, 
$83$ have heliocentric velocities within $3\sigma$ of the average velocity, 
and thus are designated as members of Leo~II.
B07 estimated the metallicities of their sample relating the equivalent widths of the CaT.

It has to be noted that K207 and B07 data sets do not overlap.

In addition, \cite{spencer:17} (hereafter S17) obtained a large data set of 
RGB member candidates of Leo~II, which were separately analyzed.
S17 performed spectroscopic observations with the Multiple Mirror Telescope 
using Hectochelle, a multifiber, single-order echelle spectrograph.
They obtained simultaneous estimates of radial velocity, effective temperature, 
surface gravity, and metallicity by fitting a library of smoothed, synthetic stellar
spectra to each Hectochelle spectrum in pixel space \citep{walker:15}.
In order to separate stellar members from nonmembers they employed a velocity 
cut. Again, stars with radial velocities larger than $3\sigma$ were taken to be
Galactic foreground stars. After performing the kinematic cut, they still  
had $11$ stars with velocities and positions similar to those of Leo~II stars. 
Therefore, they applied an extra cut in the data based on stellar surface gravities. 
From both criteria, they end with a total of $175$ Leo ~II members.

We look for stellar substructures in the Leo~II dSph, combining the RGB star
data of K207 and B07. 
We applied the {\em brute-force} method setting the ranges for the 
metallicity $met_{i}=-3.4$~dex, $met_{f}=-1.$~dex and $\Delta met=0.3$~dex. For 
the velocities we set $v_{i}=-20$~km~s$^{-1}$ and $v_{f}=20$~km~s$^{-1}$, with 
$\Delta v=4$~km~s$^{-1}$. The typical measurement errors for the metallicity and 
the velocity (as in Leo~I) are $0.11$~dex and $5.5$~km~s$^{-1}$, respectively.

Using the {\em brute-force} algorithm and the MST method, we did not find any 
significant stellar substructure in the K207 + B07 data set. 
The non-detection of stellar substructures in these data might be related to the 
small sample size ($128$ with K207 and B07 combined), and/or to the large 
Galactocentric distance of Leo~II. 

On the other hand, analyzing the S17 data with the {\em brute-force} algorithm we 
found two considerable groups of stars. In Figure~\ref{fig:LeoII_ALL} we show K207  
combined with B07 data as open circles; and K17 data as plus symbols. The two 
groups of stars are shown as filled cyan squares (hereafter \textsc{lii1}), and 
filled purple squares (hereafter \textsc{lii2}).

The substructure \textsc{lii1} is the most metal-poor group that we find. Five 
stars make up \textsc{lii1}. It has a mean metallicity of $-2.35$, a mean 
velocity of $72.4$~km~s$^{-1}$, and a velocity dispersion of $1.4$~km~s$^{-1}$, in
a elliptical annulus width of 5~arcmin.
The substructure \textsc{lii2} is constituted of seven stars. The mean metallicity 
is $-2$, the mean velocity is $78.8$~km~s$^{-1}$, and it has a velocity dispersion of 
$1.7$~km~s$^{-1}$, in a elliptical annuli width of 6~arcmin.


We compute the velocity dispersion of all the stars in the S17 data belonging 
to the same annulus as \textsc{LII1} and \textsc{LII2}. We found that the 
velocity dispersion of the elliptical annulus associated with \textsc{LII1} is 
7~km~s$^{-1}$, and the velocity dispersion associated with \textsc{LII2} is
7.6~km~s$^{-1}$.

\begin{figure*}[t]
\epsscale{1.2}
\plotone{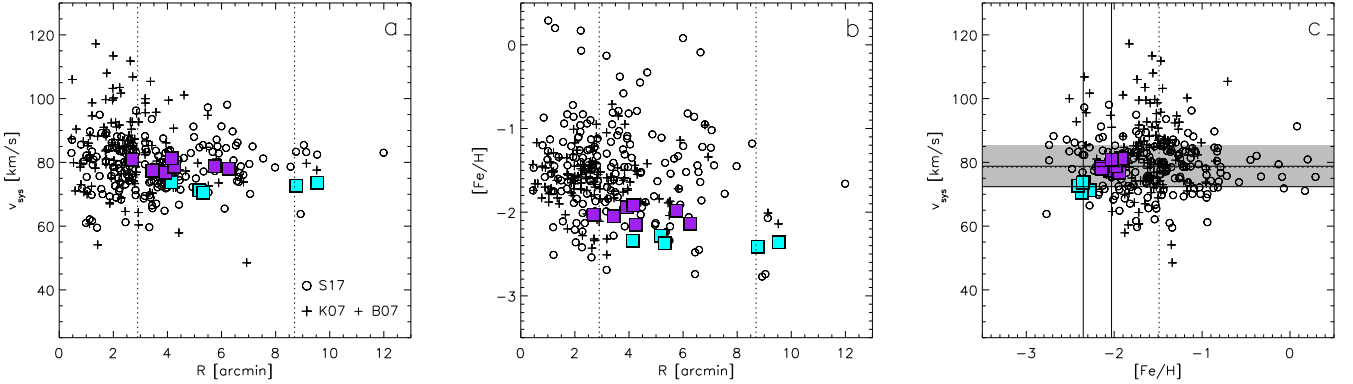}
\caption{Combined Leo~II data of K07 and B07 as black plus 
symbols. The data of S17 shown as open black circles. The two substructures found 
are plotted as filled cyan squares (\textsc{lii1}), and filled purple squares (\textsc{lii2}). 
In panel (a) we show the radial velocity as a function of the elliptical radius. In 
panel (b) we show the metallicity as a function of the elliptical radius. The 
vertical dashed lines in panels (a) and (b) indicate the core and tidal radius of Leo~II. 
In panel (c), we show the radial velocity as a function of the metallicity. The two vertical/horizontal solid
lines represent the mean metallicity/velocity for each of the substructures. The dashed vertical and horizontal
lines indicate the mean value of the metallicity and velocity of S17 sample. The shaded zone indicates the 
$1\sigma$ velocity region.}
\label{fig:LeoII_v_Fe}
\end{figure*}

In Figure~\ref{fig:LeoII_v_Fe}a we show the systemic velocity as a function of the
elliptical radius, where we can clearly observe that the substructures 
appear elongated (about $4$ to $6$ arcmin). The average velocity of \textsc{lii1} 
is very similar to the systemic velocity of Leo~II ($\delta=0.3$~km~s$^{-1}$). 
Figure~\ref{fig:LeoII_v_Fe}b shows the metallicity as a function of the elliptical 
radius. 25\% of the stars in the S17 data have metallicities lower than $\sim-2$.
Moreover, only 8.6\% of the stars in the S17 sample have metallicities lower than
$\sim -2.3$, and one third of those stars are contained in \textsc{lii1}.

The velocity as a function of metallicity is shown in Figure~\ref{fig:LeoII_v_Fe}c.
In the velocity-metallicity space, one can clearly observe both substructures clumped 
together. The mean velocity of \textsc{lii1} is $1\sigma$ away from Leo~II's average 
velocity (shown as a shaded region in Figure~\ref{fig:LeoII_v_Fe}c). The low metallicity
of \textsc{lii1} together with its velocity makes it a thought-provoking stellar 
substructure.

It is worthwhile mentioning that \cite{komiyama:07} carried out wide-field $V,I$ 
imaging of Leo~II extending far beyond Leo~II's tidal radius. They reported the
existence of a substructure in the eastern part of the galaxy (containing four 
bright RGB stars) with a physical size of $270\times170$~pc$^2$ located beyond 
the tidal radius ($8.63'$), with a luminosity close to that of a globular cluster (see 
black star in Figure~{\ref{fig:LeoII_ALL}}). They suggest that this substructure 
is a disrupted globular cluster that is merging with the main stellar component 
of Leo~II.

Interestingly, two of the stars in \textsc{lii1} are located beyond the tidal 
radius, and the interpretation could be similar to that of the knotty stellar 
structure reported by \cite{komiyama:07}. We computed a first order approximation 
on the stellar mass of \textsc{lii1} making the assumption that the number of RGB stars
in the S17 data is representative of Leo~II. Then, \textsc{lii1} would account
for $\sim 3\%$ of Leo~II's total stellar mass. Taking a luminosity of 
$L_{V}=7.4\times10^{5}$~L$_{\odot}$ \citep{coleman:07} and a typical value
(for globular clusters) of the mass-to-light ratio equal to two, then 
\textsc{lii1}'s mass would be $\sim4.4\times10^{4}$~M$_{\odot}$.

We perform a statistical study to investigate whether random groupings of 
stars in the S17 data are rare. We carried out Monte Carlo tests of normal 
distributions for both metallicity and velocity (see 
Section~\ref{sec:methods:brutus}). We find that the probability of finding 
\textsc{LII1} is $\sim5\%$ (i.e., $2.1\sigma$).
On the other hand, the probability of finding \textsc{LII2} is only $1.27\sigma$. 
Therefore, we conclude that \textsc{LII2} is not statistically significant.
Indeed, \textsc{LII1} could be real stellar debris, but we have to keep in mind 
that we are dealing with a low number of Leo~II members.

One must keep in mind that a high fraction of binary stars could falsify 
radial velocity measurements. Recently, \cite{spencer:17b} determined a binary fraction 
of $0.3-0.34$ for the Leo~II dSph galaxy. 


%
%
%

\section{Conclusions}
\label{sec:conclusions}

%
%
In this paper we searched for stellar substructures in four dSph galaxies that are satellites of 
the MW. We were able to find the stellar substructure reported by \cite{battaglia:11}, 
and a new substructure in Sextans, \textsc{sxt2}. The latter stellar substructure consists 
of eight stars with metallicities from $-2.99$ to $-2.63$~dex. Moreover, if we relax 
the constraint on the velocity ($3.7-16$~km~s$^{-1}$) we can add two more stars to the clump 
maintaining the same low metallicity range. The distance from the center of Sextans to 
the center of the ten-star clump is $\sim751$~pc, and the velocity dispersion is cold
($\sigma\simeq2.01$~km~s$^{-1}$). If the stars of the 
\textsc{sxt2} clump belong to a disrupted globular cluster then these low metallicities would 
suggest that it would be one of the most metal-poor globular clusters 
known. It is very encouraging to see that \textsc{sxt2} lies in the densest region reported
by \cite{roderick:16}.

We also find a cold stellar substructure close to the core ($240$~pc) of the Carina 
dSph. This substructure \textsc{car1}, consists of nine stars with metallicities ranging from 
$-2.89$ to $-2.44$~dex and a velocity dispersion of $\sigma\simeq1.88$~km~s$^{-1}$. 
Such a substructure resembles a disrupted globular cluster, 
very similar to that found by \cite{battaglia:11} in Sextans (\textsc{sxt1}). It has 
to be noted that the distance from the center of Carina to the center of 
Carina's substructure is only $\sim14.4$~pc, i.e., very close to the center of Carina. 
Interestingly, the \textsc{car1} substructure could possibly be related to the 
substructure reported by \cite{fabrizio:11}, since both lie close 
to the core of Carina.

Analyzing Leo~I, we find a new substructure, besides the one reported by K07 and
M08. The \textsc{li1} substructure has six stars with metallicities ranging from 
$-1.78$ to $-1.52$~dex, an average velocity of $286$~km~s$^{-1}$, and a velocity 
dispersion of $\sigma\simeq1.6$~km~s$^{-1}$. After the statistical analysis, we
concluded that \textsc{li1} is not significant.

In the dSph galaxy Leo~II, we found two significant groups of stars, \textsc{lii1}
and \textsc{lii2}. From those two groups, we found that only \textsc{lii1} is
statistically significant. Even if the probability of finding \textsc{lii1} is 
very low ($0.5$\%), we have to mention that we are dealing with low number 
statistics.

The finding of new stellar substructures is of major importance, since it leads 
to the idea that the early evolutionary histories of dSph are as complicated as 
the ones of massive galaxies.
In particular, stellar substructures are of great use when investigating the 
DM profiles of dwarf galaxies. 
For example, using $N$-body simulations, one can study the survival of
old cold substructures in the DM halo of dSphs against phase mixing,
and compare the evolution of the stellar substructures when the dark halo has 
a core and when the dark halo has a cuspy profile \citep{kleyna:03, lora:09, 
lora:12, lora:13, contenta:17, amorisco:17}.
The existence of the old stellar clump in UMi \citep{kleyna:98} is in agreement with 
a cored DM profile rather than a cuspy NFW one. If the old stellar clump in 
UMi is dropped in a NFW cuspy profile, the stellar clump gets disrupted in the first 
Gyr \citep{kleyna:03,lora:09,lora:12,lora:13}. 

\cite{lora:13} also found that a cored DM profile is needed
in order to guarantee the survival of the old stellar substructure found in Sextans 
by \cite{battaglia:11}, and the one found by \cite{walker:06}.
Then, the new stellar substructures found in this work will be crucial to further
investigate the core/cusp problem in other dSphs.
Therefore, we will perform N-body simulations of the four dSph studied in this paper 
and the new stellar substructures to investigate their evolution within their
parent DM halo.

Given the very small number of stars found to be associated with the substructures
in the current study and in others, and given how difficult it is to destroy GCs
in the absence of strong tidal fields, one might also suggest that we are looking 
here at the possible remnants of very old, metal-poor, low-mass clusters akin to
open clusters.
Either way, another interesting question is how such clusters were able to form to
begin with in systems that we see nowadays as very low stellar density objects.
Moreover, one could argue that these discoveries may add to the recent discoveries
of old clusters in low-mass dSphs, making them a more common occurrence than previously 
thought.

It is worthwhile to mention that \cite{amorisco:14} point out that mergers 
of low-mass galaxies are expected within the hierarchical model of galaxy formation. 
Moreover, they report the kinematic detection of a stellar stream in the dSph 
Andromeda~II, which they suggest could be the remnant of a merger between two dwarf 
galaxies (see also \citeauthor{koch:12} \citeyear{koch:12} for low-mass range). 
Thus, further study on the properties of the stellar substructures 
reported in this work might shed light on the way galaxies assemble through mergers at 
very small scales.



\acknowledgments
V.L. gratefully acknowledges support from the \mbox{CONACyT} Research Fellowships program. 
V.L. thanks Giuseppina Battaglia, Mathew Walker and Yutaka Komiyama for making their data 
available.
V.L. thanks Alejandro Raga, Gustavo Bruzual and Sundar Srinivasan for very helpful comments, 
suggestions and discussions which resulted in an improved version of this paper.

S.S. was supported by Sonderforschungsbereich SFB 881 ``The Milky Way System'' (subproject B5,
funding period 2011-2014) of the German Research Foundation (DFG) during part of this work.

EKG and AK were supported  by Sonderforschungsbereich SFB 881 "The Milky Way System" 
(subprojects A02 and A08) of the German Research Foundation (DFG).

We thank the anonymous referee for very kind and useful comments that improved the presentation 
of this paper.




\clearpage
%

%

\end{document}